\documentclass[aps,twocolumn,floatfix,amsmath,amssymb]{revtex4}
\usepackage{graphicx}
\usepackage{dcolumn}
\usepackage{bm}
\usepackage{epstopdf}
\usepackage[]{SIunits}
\usepackage{units}
\usepackage[colorlinks, linkcolor = blue, citecolor = blue, filecolor = black, urlcolor = blue]{hyperref}

\newcommand{\kb}[1]{\text{\tiny{#1}}}

\newcommand{\kd}[1]{\mathbf{#1}}

\begin{document}

\title{Alignment of D-state Rydberg molecules}
\author{A.T. Krupp}
\affiliation{5. Physikalisches Institut, Universit\"{a}t Stuttgart,Pfaffenwaldring 57, 70569 Stuttgart, Germany}
\author{A. Gaj}
\affiliation{5. Physikalisches Institut, Universit\"{a}t Stuttgart,Pfaffenwaldring 57, 70569 Stuttgart, Germany}
\author{J.B. Balewski}
\affiliation{5. Physikalisches Institut, Universit\"{a}t Stuttgart,Pfaffenwaldring 57, 70569 Stuttgart, Germany}
\author{P. Ilzh\"{o}fer}
\affiliation{5. Physikalisches Institut, Universit\"{a}t Stuttgart,Pfaffenwaldring 57, 70569 Stuttgart, Germany}
\author{S. Hofferberth}
\affiliation{5. Physikalisches Institut, Universit\"{a}t Stuttgart,Pfaffenwaldring 57, 70569 Stuttgart, Germany}
\author{R. L\"{o}w}
\affiliation{5. Physikalisches Institut, Universit\"{a}t Stuttgart,Pfaffenwaldring 57, 70569 Stuttgart, Germany}
\author{T. Pfau}
\affiliation{5. Physikalisches Institut, Universit\"{a}t Stuttgart,Pfaffenwaldring 57, 70569 Stuttgart, Germany}
\author{M. Kurz}
\affiliation{Zentrum f\"{u}r Optische Quantentechnologien, Universit\"{a}t
Hamburg, Luruper Chaussee 149, 22761 Hamburg, Germany}
\author{P. Schmelcher}
\affiliation{Zentrum f\"{u}r Optische Quantentechnologien, Universit\"{a}t Hamburg, Luruper Chaussee 149, 22761 Hamburg, Germany}
\affiliation{Hamburg Centre for Ultrafast Imaging, Universit\"{a}t Hamburg,
Luruper Chaussee 149, 22761 Hamburg, Germany}

\date{\today}


\begin{abstract}

We report on the formation of ultralong-range Rydberg D-state molecules via
photoassociation in an ultracold cloud of rubidium atoms. By applying a magnetic
offset field on the order of \unit[10]{G} and high resolution spectroscopy, we
are able to resolve individual rovibrational molecular states. A full theory, using the Born-Oppenheimer approximation 
including s- and p-wave scattering, reproduces the measured
 binding energies. The calculated molecular wavefunctions show that in the
 experiment we can selectively excite stationary molecular states
 with an extraordinary degree of alignment or anti-alignment with respect to the magnetic field axis.
  
\end{abstract}

\maketitle

%
Angular confinement of molecules, referred to as alignment, represents
a unique way of influencing molecular motions. It is of major importance
for the control of a number of molecular processes and properties, such as the pathways of chemical reactions including
stereo-chemistry ~\cite{brooks:jcp45, Stolte:BBGPC86:413, zare:science, aquilanti:pccp_7},
photoelectron angular distributions~\cite{Holmegaard:natphys6,PhysRevA.83.023406,Landers:PRL87:013002,bisgaard:science323},
dissociation of molecules~\cite{wu:jcp101,baumfalk:jcp114,brom:11645,lipciuc:123103}
and diffractive imaging of molecules~\cite{PhysRevLett.92.198102,Filsinger:PCCP13:2076}.
In the case of ultracold alkali dimers, the quantum stereodynamics of ultracold
bimolecular reactions has been probed recently \cite{Miranda:NatPhys}.
To achieve alignment and its ally orientation
electric, magnetic and light fields, have been used in a variety of experimental configurations
such as, e.g., the brute force orientation~\cite{Loesch1990}, hexapole
focusing~\cite{brooks:science,parker:1989,Hain1999}, strong ac pulsed fields
\cite{Seide2003} or combined ac and dc electric
fields~\cite{friedrich:jcp111,friedrich:jpca103,baumfalk:jcp114, sakai:prl_90, Tanji2005}.
They all have in common that they provide an angular-dependent potential
energy that leads to a hybridisation of the field-free rotational motion.
Beyond the above it is well-known that in strong magnetic fields the mutual orientation
of the magnetic field and internuclear axis provides an intricate electronic state-dependent
topology of the corresponding adiabatic potential energy surfaces (APES) yielding a plethora
of equilibrium positions \cite{Kappes}, novel bonding mechanisms
\cite{Detmer,Helgaker2012} and field-induced vibronic interactions via e.g.
conical intersections of the APES \cite{Kappes,Schmelcher1990}.\\
In the present work we show that weak magnetic fields of a dozen Gauss allow to
strongly impact and control the properties of ultralong-range Rydberg molecules. Rydberg molecules have been
theoretically predicted \cite{Greene2000,Lesanovsky2006} and experimentally
observed for Rydberg S-states \cite{Bendkowsky2009,Bendkowsky2010} and P-states
\cite{Bellos2013}.\\

Here we investigate D-state ultralong-range
rubidium Rydberg molecules for two different $m_J$
magnetic substates with high resolution spectroscopy. We can
selectively excite distinct rovibrational molecular states with specific
alignments and identify them by comparison of the binding energy with
theoretical predictions.
\begin{figure*}[!htb]
\includegraphics{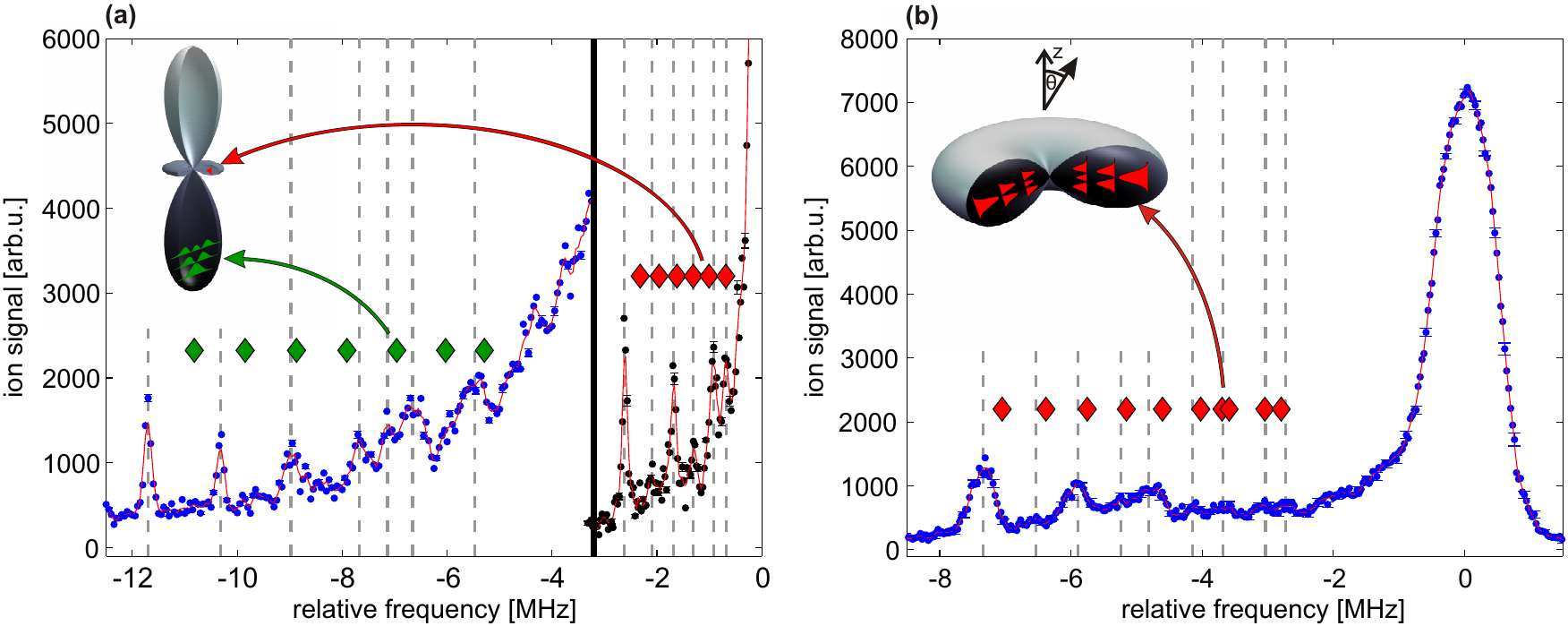}
\caption{\label{fig1} Spectra of the 44D,
\textit{J}=5/2, $m_J$=1/2 (a) and 42D, \textit{J}=5/2, $m_J$=5/2 (b) states,
where the ion detector signal is plotted against the relative frequency to the
atomic line. In (a) the spectrum consists of two individual spectra taken with
different laser intensities separated with a black line at \unit[-3.2]{MHz}:
the left one (blue) was taken at a high intensity to resolve the axial
molecules and the right one (black) at a low intensity to decrease the power broadening of
the atomic line such that we can resolve the toroidal lying molecules. For
better visibility a moving average (red line) is included. To make it easier to identify the molecular positions the data was
scaled by a factor of 3 for the $m_J$=1/2 state. The dashed lines (grey) mark
the experimental peak positions of the molecules
whereas the red and green diamonds indicate the calculated binding energies; 
the green diamonds were used for the axial molecules in case of
$m_J$=1/2. The insets show the angular part of the electron orbitals relevant
for triplet scattering. Arrows point to the positions where the
molecules are created within the orbitals. The standard deviation errorbars are
determined from independent measurements.}
\end{figure*}
%
The bond of the ultralong-range Rydberg molecules results from the
low-energy scattering between a quasi-free Rydberg electron at position
\textbf{r} and a ground state atom at \textbf{R}. This process can be
described using the Fermi pseudopotential \cite{Fermi1934},
which depends on the scattering length $A$ between two scattering
partners:
\begin{equation}
\begin{array}{rl}
V_{\textrm{n,e}}(\bf{\textbf{r}},\bf{\textbf{R}})=&2\pi A_{s}[k(\textbf{R})]\delta(\bf{\textbf{r}}-\bf{\textbf{R}}) \\
&+6\pi A^{3}_{p}[k(\textbf{R})]\overleftarrow{\nabla} \delta(\bf{r}-\bf{R})
\overrightarrow{\nabla},\label{potential}
\end{array}%
\end{equation} 
where $A_{s}(k)$ and $A_{p}(k)$ are
the 
s-wave and p-wave triplet scattering
lengths, respectively \cite{Omont1977}. The momentum $k$ of the electron can be
treated in a semiclassical approximation \cite{Greene2000}.
The resulting Hamiltonian 
\begin{eqnarray}
H = H_0 + \frac{B}{2}(J_{z}+S_z)+V_{\rm{n,e}}(\bf{r},\bf{R})+
\frac{{\bf{P}}^2}{\textit{M}}
\label{hamiltonian}
\end{eqnarray}
consists of the field-free Hamiltonian $H_0$ of the Rydberg atom, the Zeeman-interaction
terms of the angular momenta (spin \textbf{S} and orbital \textbf{L}) with
the external field $\bf{B}=$$B$$\bf{e}_{\rm{z}}$, the scattering potential
\eqref{potential} and the kinetic energy term. Here the total angular
momentum $\bf{J}=\bf{L}+\bf{S}$ was introduced.
We write the total wavefunction
as $\Psi(\bf{r},\bf{R})=\psi(\bf{r}, \bf{R} )\phi(\bf{R})$, where
$\psi$ describes the electronic molecular wavefunction
in the presence of the neutral perturber for a given position $\bf{R}$ and $\phi$ determines
the rovibrational state of the perturber. The resulting APES
$\epsilon(R,\Theta)$ depend on the angle of inclination $\Theta$ between the field vector and the
internuclear axis as well as the internuclear distance $R$. By solving the 
Schr{\"o}dinger equation in cylindrical coordinates using a finite difference
method we obtain the binding energies and molecular wavefunctions without
any fitting parameters (see Supplementary Material).\\
In the experiment we start with an ultracold (2 \unit{\micro\kelvin}) cloud
of about 5$\cdot10^5$ $^{87}$Rb atoms in a magnetic trap (peak density 
$\sim$10$^{13}$cm$^{-3}$) polarized in the  
5S$_{1/2}$, $F$=2, $m_F$=2 state.
For the photoassociation of the molecules, a
$\sigma^+$-polarized laser at \unit[780]{nm}, \unit[500]{MHz} detuned
from the intermediate $5P_{3/2}$ state, and a laser at \unit[480]{nm}
(combined laser linewidth $<$ \unit[30]{kHz}) are used. A magnetic field
of $B$=\unit[13.55]{G} is applied to separate the different
atomic $m_J$ states, leading to a Zeeman splitting of the fine structure states
of $\sim$\unit[22]{MHz}. After the \unit[50]{\micro\second} long Rydberg
excitation pulse we field-ionize the Rydberg states and accelerate the ions towards a microchannel plate detector. In a single cloud we
perform up to 400 cycles of excitation and detection while scanning
the laser frequency.  This permits us to take one spectrum within a minute.
More information about the experimental setup can be found in
\cite{LWN12}.
We investigate the stretched state \textit{J}=5/2, $m_J$=5/2 and the
\textit{J}=5/2, $m_J$=1/2 state. To address only these states we change the
polarization of the \unit[480]{nm} laser to either $\sigma^+$ ($m_J$=5/2)
or $\sigma^-$ ($m_J$=1/2).
For principal quantum numbers~\textit{n}
ranging from 41 to 49 the total angular momentum
quantum number \textit{J} is still a good quantum number since the fine
structure splitting of $\sim$ 170 to \unit[98]{MHz} (for \textit{n}=41 to
49) is large compared to the Zeeman splitting.
This region was chosen as for lower quantum numbers $n<40$ the binding 
energies of the outermost molecular states
are on the same energy scale as the Zeeman splitting. This would lead
to an undesired overlap of the molecular states with the neighboring atomic
line.
For higher principal quantum numbers $n>50$ the distance between
neighboring molecular lines decreases below our spectral resolution
\cite{Balewski2013}. \\
%
%
%
\begin{figure}[t]
\includegraphics[width=\columnwidth]{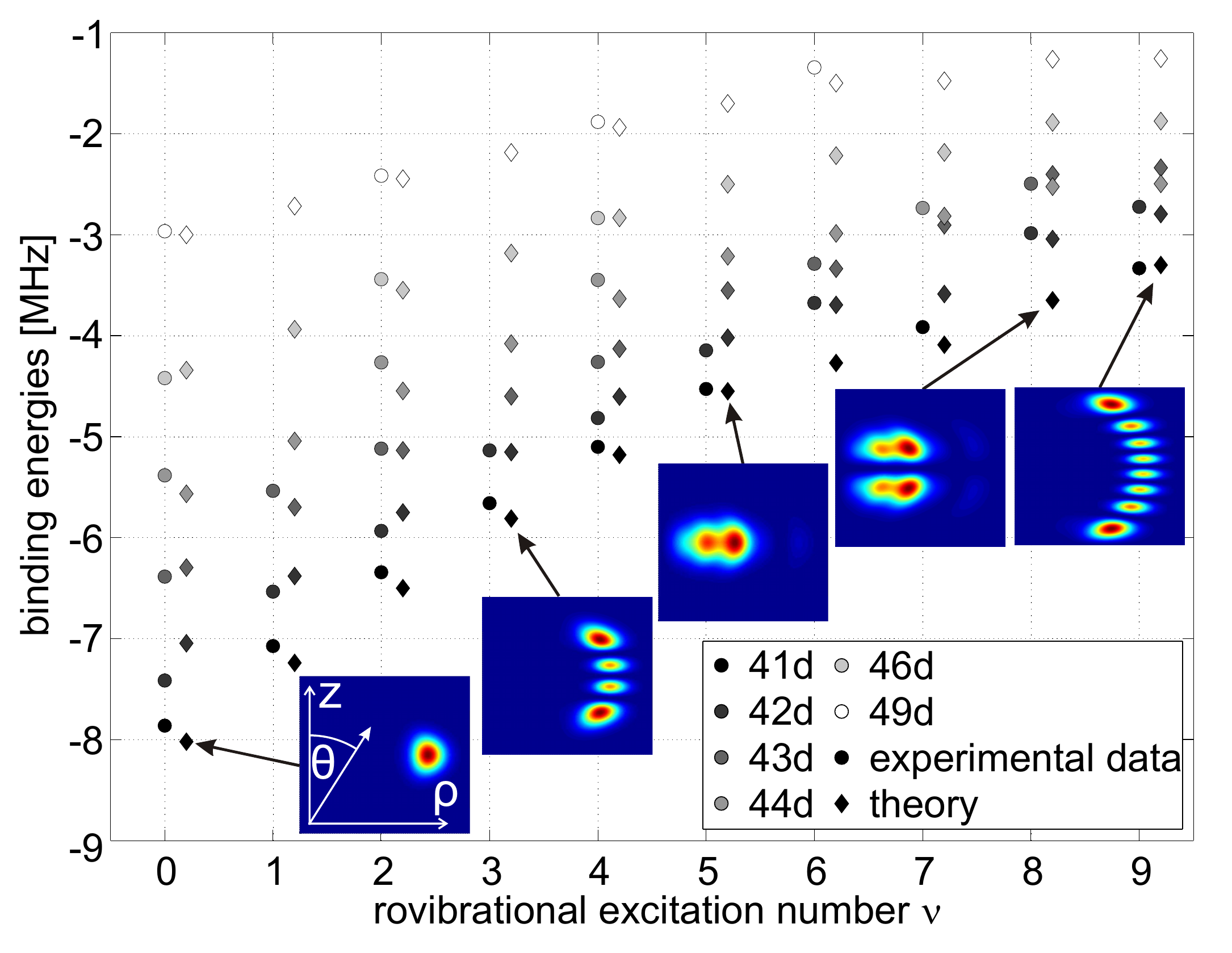}
\caption{\label{fig2} Molecular binding energies for the $m_J$=5/2 states  
plotted against the rovibrational excitation numbers $\nu$ for principal quantum
numbers $n$ ranging from 41 to 49. For increasing $n$ the states are colored brighter. The
calculated binding energies (diamonds) are plotted with a horizontal
offset to the experimental ones (circles) to improve readability. The insets
depict the probability densities ranging from $\rho$=2000$a_0$ to
$\rho$=3300$a_0$ and for $z$=-1500$a_0$ to $z$=1500$a_0$ of certain
rovibrational states.}
\end{figure}
\begin{figure*}
\includegraphics{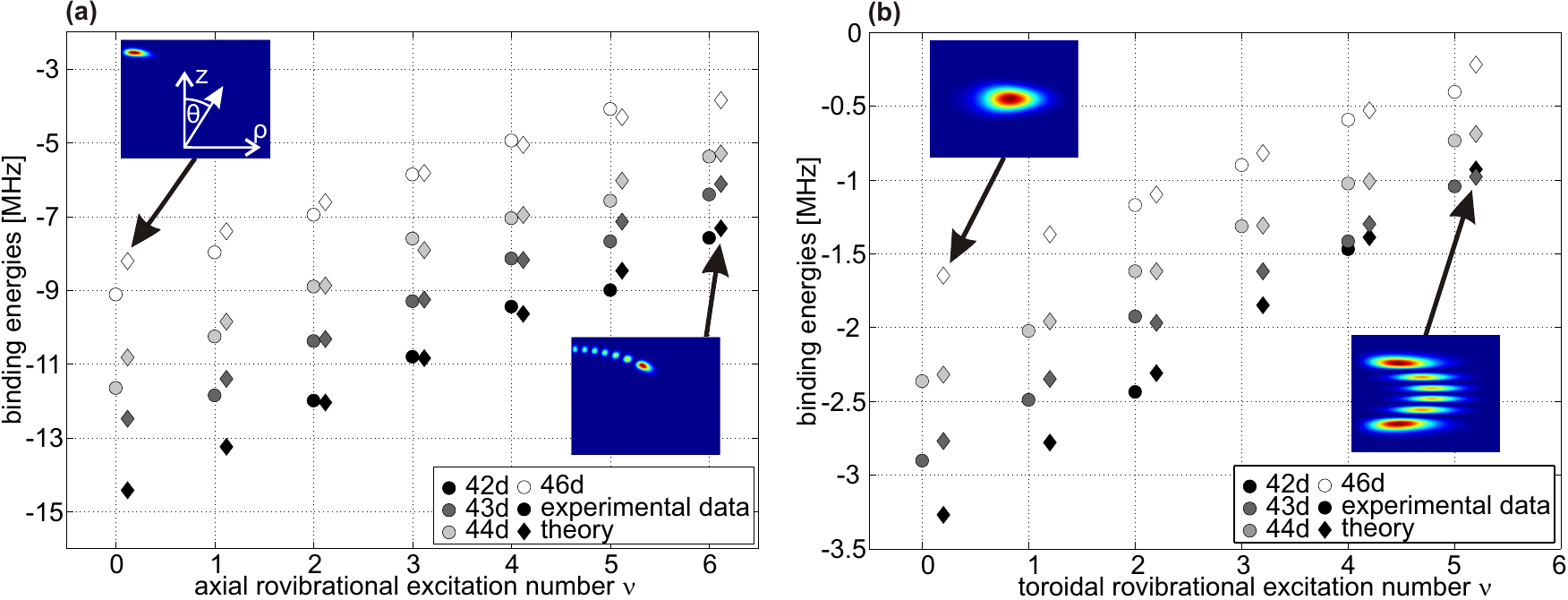}
\caption{\label{fig3} Molecular binding energies for the $m_J$=1/2 states
plotted against the axial rovibrational excitation numbers (a) and the toroidal
rovibrational excitation numbers (b), respectively, for principal quantum numbers $n$
ranging from 42 to 46. For increasing $n$ the states are colored brighter. The
calculated binding energies (diamonds) are plotted with a horizontal
offset to the experimental ones (circles) to improve readability. The insets depict the probability densities in $z$- and
$\rho$-direction of selected rovibrational states. For the axial rovibrational
excitation numbers (a) $\rho$ ranges from 0 to 3500$a_0$ and
$z$ from 0 to 3500$a_0$ whereas the toroidal rovibrational excitation
numbers range from (b) $\rho$=2700$a_0$ to 3500$a_0$ and $z$=-2250$a_0$ to
2250$a_0$ (further details can be found in the supplementary material).}
%
%
%
\end{figure*}
%
%
%
%
%
%
Photoassociation spectra of the 44D, \textit{J}=5/2, $m_J$=1/2 and 42D,
\textit{J}=5/2, $m_J$=5/2 states are shown in Fig.\ref{fig1} (a) and (b),
respectively.
The confinement of the electron density in the polar coordinate~$\Theta$ leads
to a large number of excited rovibrational states visible in the spectrum in contrast
to previous S-state measurements \cite{Bendkowsky2009,Bendkowsky2010}.
This causes stationary molecular states featuring different degrees of
alignment. \\
For the $m_J$=5/2 state (Fig.\ref{fig1} (b)), the molecular states are
anti-aligned in a plane perpendicular to the quantization axis, at $\Theta=\pi/2$.
The molecular potential at this position is 
$|Y_{l=2}^{m=2}(\Theta=\pi/2)|^2/|Y_{l=0}^{m=0}|^2=1.875$ times deeper
than in the well known Rydberg S-state molecules.
This factor explains well the measured binding energies of the deepest bound
states.
In addition the energies of the excited rovibrational states are 
reproduced by our calculations indicated as red diamonds.
For the $m_J$=1/2 state two classes of 
molecular states appear, one localized in the polar lobes ($\Theta=0,\,\pi$;
green) and the other one in the toroidal part of the orbital in the equatorial
plane ($\Theta=\pi/2$; red).
The angular part of the molecular potential of the $m_J$=1/2 state including the
Clebsch-Gordan coefficient scales as $3/5\cdot|Y_{l=2}^{m=0}(\Theta)|^2$. Note that
this state is a superposition of a singlet and triplet state, where we can neglect
the singlet part due to its small scattering length \cite{Bahrim2001}.
As a result the lowest aligned axial molecular state shows a binding energy
four times larger than the one for the anti-aligned toroidal case and three
times larger than for the corresponding S-state molecules. Both estimates are in good
agreement with the experimental results.
The calculated binding energies of the excited states are indicated by red and
green diamonds in Fig.\ref{fig1}(a).
The difference in strength of the two classes of molecular states can be attributed to
the different spatial extent of the potential wells leading to larger Franck-Condon factors
for the anti-aligned toroidal states.
The agreement of the measured binding energies with the results of our calculations over a wide range of principal quantum numbers  
is most evident in Fig.\ref{fig2} and Fig.\ref{fig3}.
It is worth to mention that the energy of rotation and vibration are of the same order of magnitude;
thus the spectroscopic lines cannot be assigned to rotational and vibrational states separately and only one rovibrational quantum number~$\nu$ is used. 
From the volume of the Rydberg atom, one obtains a scaling of the potential
depth with the effective principal quantum number as~$n^{*-6}$.
The binding energy, however, also depends on the shape of the potential,
so that the scaling law does not describe our high resolution data
sufficiently.\\
All in all, the full calculation of the binding energies fit the
experimental $m_J$=5/2 state data (Fig.\ref{fig2}) and the $m_J$=1/2 data for the
toroidal molecules (Fig.\ref{fig3}(b)) well. 
In the insets of Fig.\ref{fig2} and Fig.\ref{fig3} the probability densities of specific
rovibrational states in $z$ and $\rho$-direction are shown. 
From these color plots, the variable degree of the alignment, defined as 
$\left<\text{cos}^2(\Theta)\right>$, becomes obvious.
Starting from the ground state $\nu$=0, the molecular wavefunction begins
to spread in $\Theta$-direction until it extends to the first radial excitation at $\nu$=6, valid for all $n$ and
$m_J$. In the case of the 42D, $m_J$=1/2 state we obtain an alignment of 0.01 of the toroidal ground state 
which increases with rovibrational excitation number. 
For the axial case we get alignments starting  
from 0.98 decreasing with increasing higher axial excitation numbers.\\
In conclusion, we report on the observation of D-state ultralong-range Rydberg molecules exposed
to magnetic fields in high resolution spectroscopy. The maximally stretched
$m_J$=5/2 and the $m_J$=1/2 Rydberg states lead, due to their different
electronic configurations, to adiabatic potential energy surfaces with different topologies. For $m_J=$5/2 the
two-dimensional potential landscape $\epsilon(R,\Theta)$ possesses a series of
local wells located at $\Theta=\pi$/2 which lead to anti-aligned
rovibrational states leaving their signatures in a series of peaks of the spectroscopic detection of the ultralong-range Rydberg molecules. 
On the contrary the $m_J =$ 1/2
potential surfaces exhibit a number of radial wells at $\Theta = 0,
\pi$ and a series of weaker potential wells for $\Theta=\pi$/2. The latter are caused by the
axial and toroidal character of the corresponding electronic configuration and lead to aligned and anti-aligned rovibrational
states. Spectroscopically these are observed as a sequence of peaks far off and close by to the
main atomic Rydberg transition, respectively. A change of the principal quantum
number $n$ introduces only quantitative changes to the above picture where theory and experiment show
a good agreement.
This work opens the doorway to the control of Rydberg molecular structures and
even chemical reaction dynamics by external fields. For polyatomic states, i.e.
several neutral perturbers, it can be conjectured that magnetic and/or electric fields can be used to strongly change the molecular geometry for weak field strengths which is otherwise impossible 
both for ground state molecules and also for the traditional
molecular Rydberg states containing a tight molecular positively charged core. Even more, the design
of conical intersections \cite{Schmelcher1990,Mayle2012} yielding ultrafast
decay or predissociation processes along selected chemical reaction coordinates comes into the reach of experimental progress 
in the field of ultracold molecular physics. \\
%
%
%
%
%
During the finalization of this manuscript we became aware of related work
\cite{Raithel}.
\begin{acknowledgments}
This work is funded by the Deutsche Forschungsgemeinschaft
(DFG) within the SFB/TRR21 and the
project PF 381/4-2. We also acknowledge support by
the ERC under contract number 267100 and from E.U.
Marie Curie program ITN-Coherence 265031. P.S. acknowledges financial support by
the Deutsche Forschungsgemeinschaft (DFG) through the excellence cluster The 
Hamburg Centre for Ultrafast Imaging - Structure, Dynamics, and Control of Matter on the Atomic Scale.
\end{acknowledgments}
%

%
%
%
%
%
\newpage
\thispagestyle{empty}
\quad  
\newpage
\onecolumngrid
\section{Supplementary Material}
\subsection{Molecular Hamiltonian in a magnetic field}
\begin{figure}[b]
\includegraphics[width=0.5\textwidth]{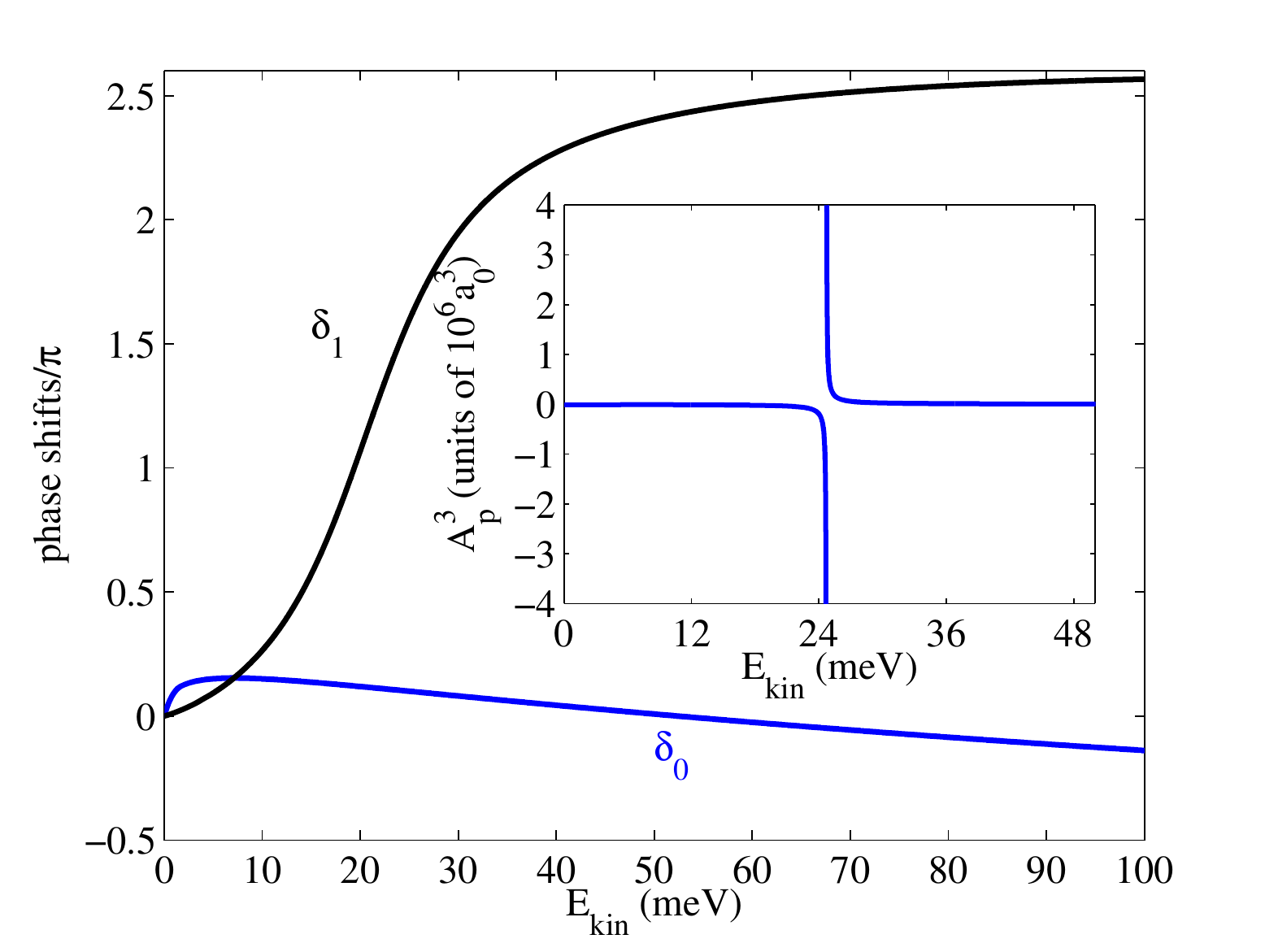}
\caption{Energy dependent triplet phase shifts $\delta_1$ and $\delta_0$ for $e^{-}-^{87}$Rb($5s$) scattering. For $E_{\rm{kin}}=24.7$meV the phase shift $\delta_1=\pi/2$, i.e.\ the (cubed) energy dependent $p$-wave scattering length $A^{3}_{p}(k)=-\tan(\delta_1(k))/k^3$ possesses a resonance at this energy. This can be clearly seen in the inset.} 
\label{fig_scat}
\end{figure}

We consider a highly excited Rydberg atom interacting with a ground state neutral perturber atom 
(we will focus on the $^{87}Rb$ atom here) in a static and homogeneous magnetic field. 
The Hamiltonian treating the $Rb$ ionic core and the neutral perturber as point particles is given by
\begin{eqnarray}
H&=&\frac{\kd{P}^{2}}{M}+H_{\rm{el}} +V_{\textrm{n,e}}(\kd{r},\kd{R}),\label{ham}\\
H_{\rm{el}}&=&H_0 + \frac{1}{2}\kd{B}(\kd{L}+2\kd{S})+\frac{1}{8}(\kd{B}\times \kd{r})^2, \label{ham_elec}
\end{eqnarray}
where $(M,\kd{P},\kd{R})$ denote the atomic $Rb$ mass and the relative momentum and position of the neutral perturber with respect
to the ionic core. The vector $\kd{r}$ indicates the relative position of the Rydberg electron to the ionic core. 
The electronic Hamiltonian $H_{\rm{el}}$ consists of the field-free Hamiltonian $H_0$ of the Rydberg atom and the usual paramagnetic and diamagnetic terms of an 
electron in a static external magnetic field. The Hamiltonian $H_0$ includes the Rydberg quantum defects due to electron-core scattering and the fine structure. $H_{\rm{el}}$ contains also the Zeeman-interaction terms of the angular momenta (spin and orbital) with the external field. We choose $\kd{B}=B\kd{e}_{\rm{z}}$. The interatomic potential $V_{\rm{n},e}$ for the low-energy scattering between the Rydberg electron and the neutral perturber is described as a Fermi-pseudopotential 
\begin{eqnarray}
V_{\textrm{n,e}}(\kd{r},\kd{R})&=&2\pi A_{s}[k(R)]\delta(\kd{r}-\kd{R})\nonumber\\
&+&6\pi  A^{3}_{p}[k(R)]\overleftarrow{\nabla} \delta(\kd{r}-\kd{R}) \overrightarrow{\nabla}.\label{inter}
\end{eqnarray}
  Here we consider the triplet scattering of the electron from the ground
  state alkali atom. $A_{s}(k)=-\tan[\delta_{0}(k)]/k$ and $A^{3}_{p}(k)=-\tan[\delta_{1}(k)]/k^3$ denote the energy-dependent triplet $s$- and $p$-wave scattering lengths. $\delta_{l=0,1}(k)$ are the energy dependent phase shifts (see Fig.\ (\ref{fig_scat})). The wave vector $k(R)$ is determined by the semiclassical relation $k(R)^2/2=E_{\rm{kin}}=-1/2n^2+1/R$ \cite{Greene2000,Omont1977}.\\
We introduce the total angular momentum $\kd{J}=\kd{L}+\kd{S}$ and write the total wave function as $\Psi(\kd{r},\kd{R})=\psi(\kd{r}; \kd{R} )\phi(\kd{R})$. Within the adiabatic approximation we obtain
\begin{eqnarray}
[H_0 + \frac{B}{2}(J_{z}+S_z)+ \frac{B^2}{8}(x^2+y^2) +V_{\kb{n,e}}(\kd{r},\kd{R})]\psi_{n,m_J}(\kd{r};\kd{R})&=&\epsilon_{n,m_J}(\kd{R})\psi_{n,m_J}(\kd{r};\kd{R}), \label{hamelec}\\
(\frac{\kd{P}^{2}}{M}+\epsilon_{n,m_J}(\kd{R}))\phi^{(n,m_J)}_{\nu m}(\kd{R})&=&E^{(n,m_J)}_{\nu m}\phi^{(n,m_J)}_{\nu m}(\kd{R}),\label{hamrovi}
\end{eqnarray}
where $\psi_{n,m_J}$ describes the electronic molecular wave function in the presence of the neutral perturber for a given 
relative position $\kd{R}$ and $\phi^{(n,m_J)}_{\nu m}$ determines the rovibrational state of the perturber.\\
\\
In this work a field strength of $B=13.55G$ is chosen. For such a field strength the diamagnetic term in (\ref{hamelec}) can be neglected. Furthermore, the adiabatic potential energy surfaces (APES) $\epsilon_{\nu m_J}(\kd{R})$ possesses rotational symmetry around the $z$-axis, which means they depend 
on the angle of inclination $\theta$ between the field vector and the internuclear axis, e. g., $\epsilon_{n,m_J}(\kd{R})=\epsilon_{n,m_J}(R,\theta)$. In case we use cylindrical coordinates, the APES are functions of $(z,\rho)$.  
\subsection{Basis set}
We calculate the APES for the $n=42,43,44,46,49,\ \ J=5/2,\ \ m_J=1/2$ and $n=41,42,43,44,46,49,\ \ J=5/2,\ \ m_J=5/2$ fine structure states. The spin orbit coupling causes a level splitting between the $J=3/2,5/2$ states in the range of $170$MHz ($n=41$) to $98$MHz ($n=49$). To obtain the potential curves we have performed an diagonalization of the electronic Hamiltonian (\ref{hamelec}) using the eigenstates $|n,J=l \pm 1/2,m_J,l=2,s=\frac{1}{2} \rangle$ of $H_0$ with
\begin{eqnarray*}
&\ &\langle \kd{r} |n,J=l \pm \frac{1}{2},m_J,2, \frac{1}{2} \rangle\\ 
&=&R_{n,j,2}(r)(\pm \sqrt{\frac{\frac{5}{2} \pm m_J}{5}}Y_{2,m_J-1/2}(\theta,\phi)|\uparrow \rangle + \sqrt{\frac{\frac{5}{2} \mp m_J}{5}}Y_{2,m_J+1/2}(\theta,\phi)|\downarrow \rangle)\\
&\equiv&R_{n,j,2}(r)(\alpha(j,m_J) Y_{2,m_J-1/2}(\theta,\phi)|\uparrow \rangle + \beta(j,m_J)Y_{2,m_J+1/2}(\theta,\phi)|\downarrow \rangle)
\end{eqnarray*}
$Y_{lm}(\theta,\phi)$ are the spherical harmonics.
\subsection{Potential energy surfaces}
We obtain APES with different topologies depending on the level of electronic
excitation (Fig.\ (\ref{fig2})-(\ref{fig3})).
The characteristic features of the $n=42,\ m_J=1/2,5/2$ potential surface which we present in Fig.\ (\ref{fig2})-(\ref{fig3}) remain up to the $n=49$ APES.
\begin{figure*}[t]
\centering
\begin{minipage}[b]{.45\textwidth}
\includegraphics[width=1.0\textwidth]{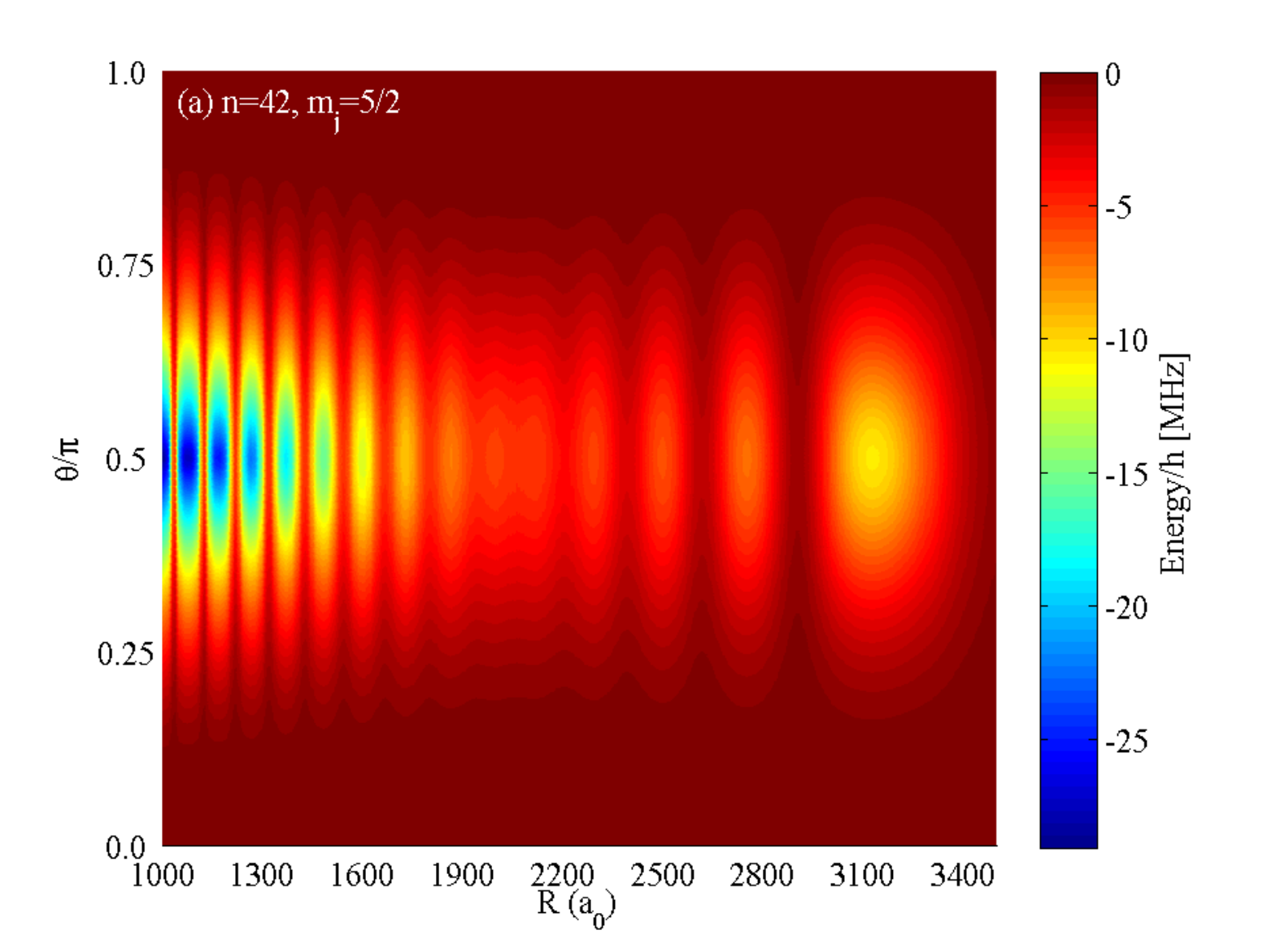}
\end{minipage}\qquad
\begin{minipage}[b]{.45\textwidth}
\includegraphics[width=1.0\textwidth]{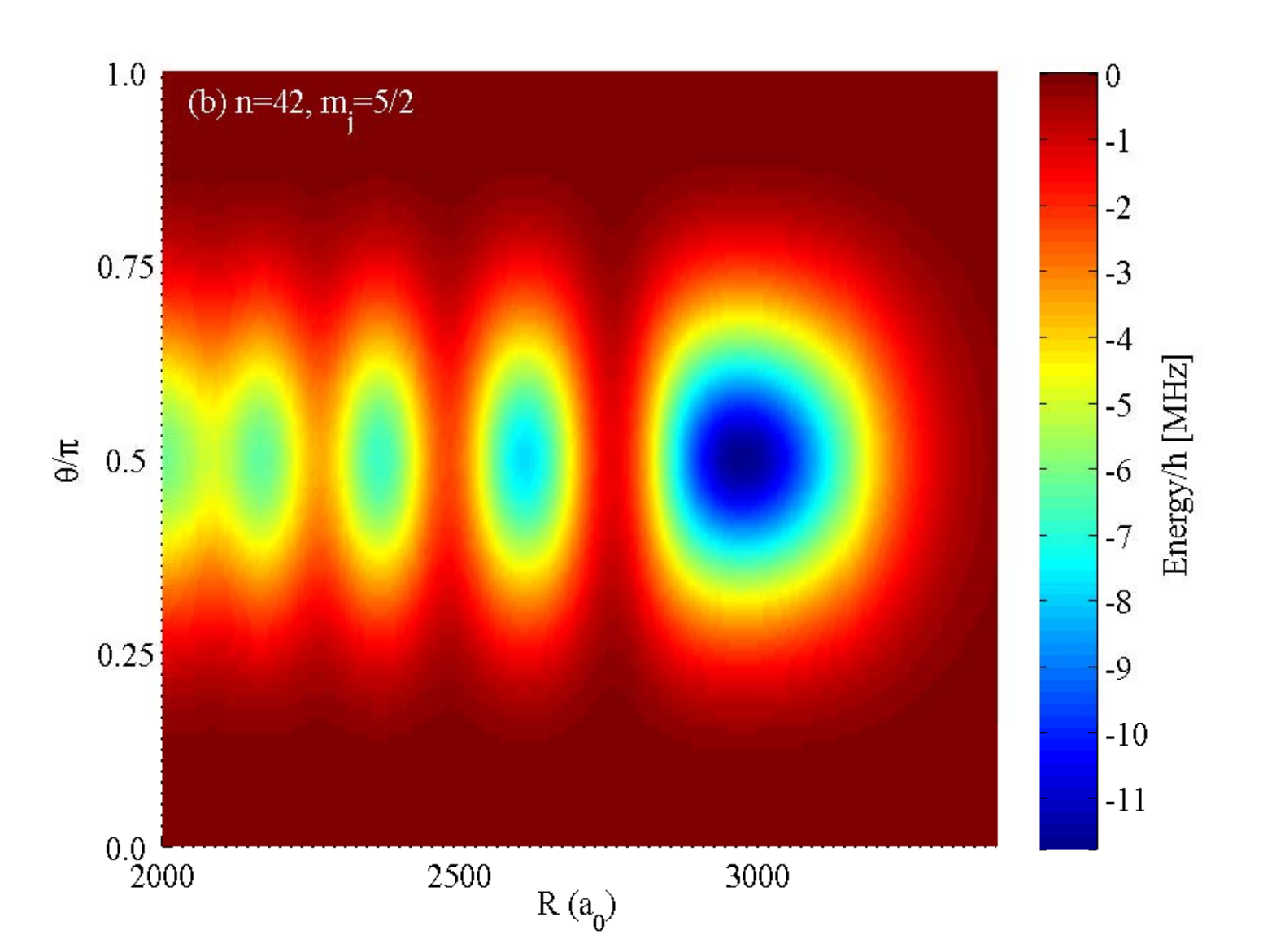}
\end{minipage}
\caption{(a) $42D_{5/2},m_J=5/2$ APES as a function of $(R,\theta)$ for $1000a_0 \le R \le 3400a_0$. One can clearly identifies a local potential minimum at $\theta=\pi/2,\ R\approx 3100a_0$ with a depth of around $12$MHz and several neighboring wells with decreasing depths. For $R \le 2000a_0$ the APES possesses a strongly oscillatory structure with a series of local potential minima increasing in depth. These oscillations are caused by the increasing impact of the $p$-wave scattering term which possesses a resonance at $R_{\rm{res}} \approx 800a_0$. Figure (b) shows the same APES but in the range $2000a_0 \le R \le 3400a_0$}\label{fig2}
\end{figure*}
\begin{figure*}[t]
\centering
\begin{minipage}[b]{.45\textwidth}
\includegraphics[width=1.0\textwidth]{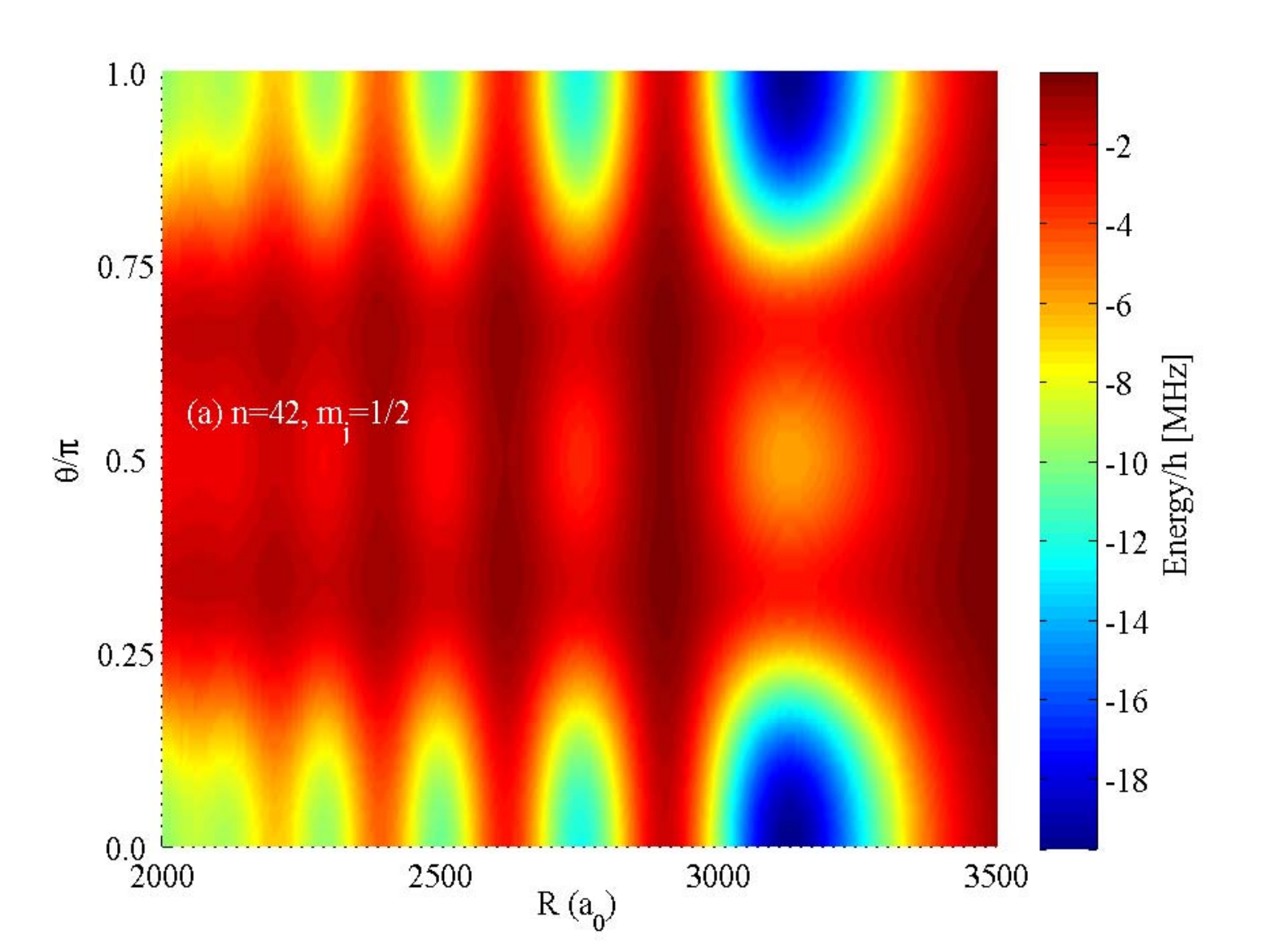}
\end{minipage}\qquad
\begin{minipage}[b]{.45\textwidth}
\includegraphics[width=1.0\textwidth]{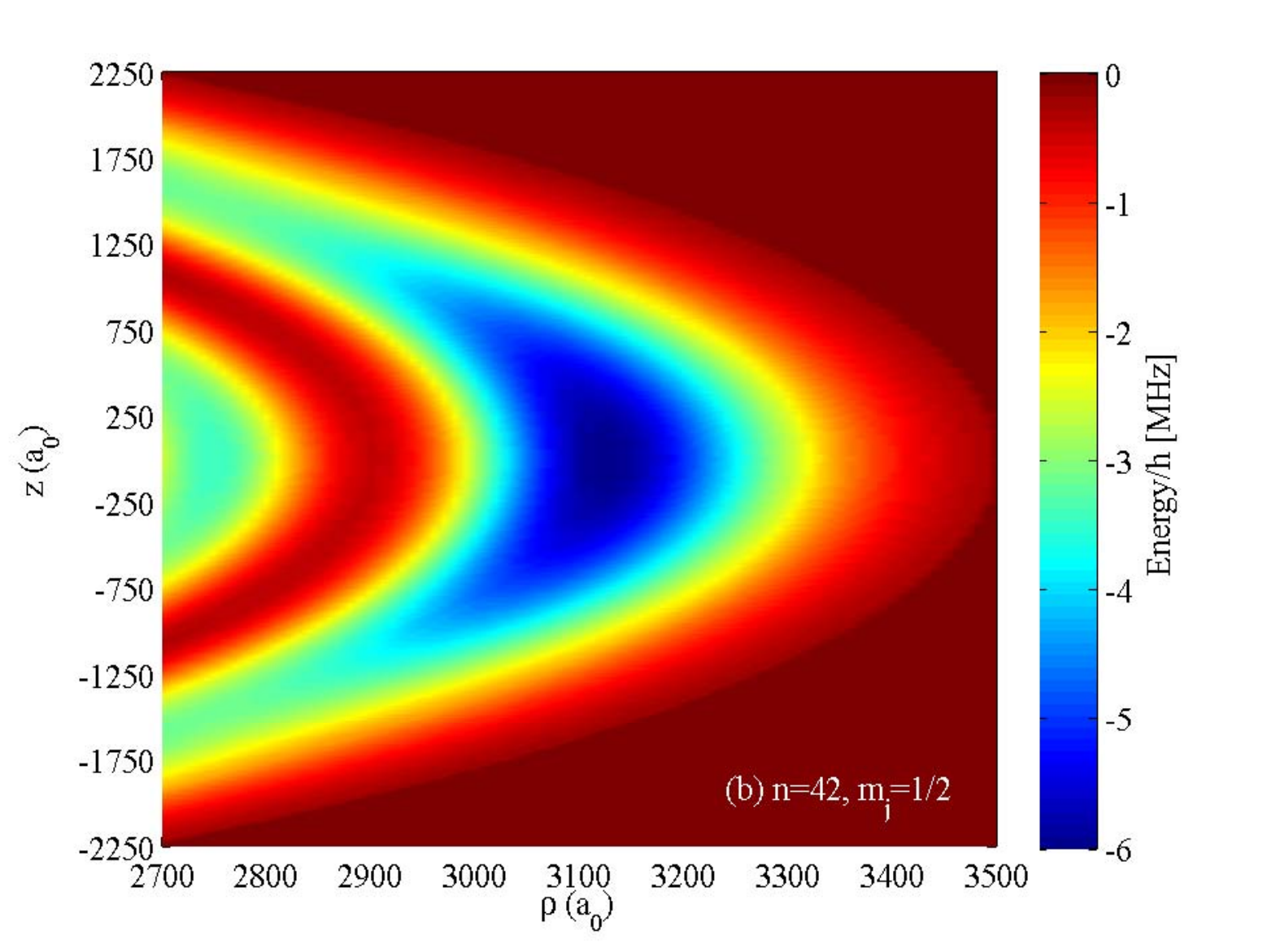}
\end{minipage}
\caption{(a) $42D_{5/2},m_J=1/2$ APES as a function of $(R,\theta)$ for $2000a_0 \le R \le 3500a_0$. We clearly see two potential wells at $\theta=0,\pi;\ R\approx 3150a_0$ with a depth of $20$MHz. In addition, a more shallow well with a depth of $6$MHz can be identified at $\theta=\pi/2,\ R\approx 3150a_0$. With decreasing $R$ neighboring potential wells decrease in depths. Figure (b) shows the vicinity of the shallow potential well in cylindrical coordinates.}\label{fig3}
\end{figure*}
\subsection{Rovibrational levels and binding energies}
For the rovibrational wavefunctions we choose the following ansatz
\begin{eqnarray}
\hspace{-0.4cm}\phi^{(n,m_J)}_{\nu m}(\kd{R})=\frac{F^{(n,m_J)}_{\nu m}(\rho,z)}{\sqrt{\rho}}\text{exp}(im \varphi),\ m \in \mathbb{Z},\ \nu \in \mathbb{N}_0 .
\end{eqnarray}
With this we can write the rovibrational Hamiltonian as 
\begin{eqnarray}    
H_{\rm{rv}}&=&-\frac{1}{M}(\partial^{2}_{\rho}+\partial^2_{z})+\frac{m^2-
1/4}{M\rho^2}+\epsilon_{n,m_J}(\rho,z).\label{sch}
\end{eqnarray}
We solve this differential equation using a finite difference method. For a
fixed $m$ we label the eigenenergies with $\nu=0,1,2,...$ and define the binding energy $E^{(\nu)}_B$ of an eigenstate as the absolute value between the eigenenergy and the dissociation limit of the APES. Because $\epsilon_{n,m_J}(\rho,-z)=\epsilon_{n,m_J}(\rho,z)$ the functions $F^{(n,m_J)}_{\nu m}$ fulfill $F^{(n,m_J)}_{\nu m}(\rho,-z)=\pm F^{(n,m_J)}_{\nu m}(\rho,z)$, which means $|F^{(n,m_J)}_{\nu m}(\rho,-z)|^2=|F^{(n,m_J)}_{\nu m}(\rho,z)|^2$ for the probability density.
\begin{figure*}[h]
\centering
\begin{minipage}[b]{.4\textwidth}
\includegraphics[width=1.0\textwidth]{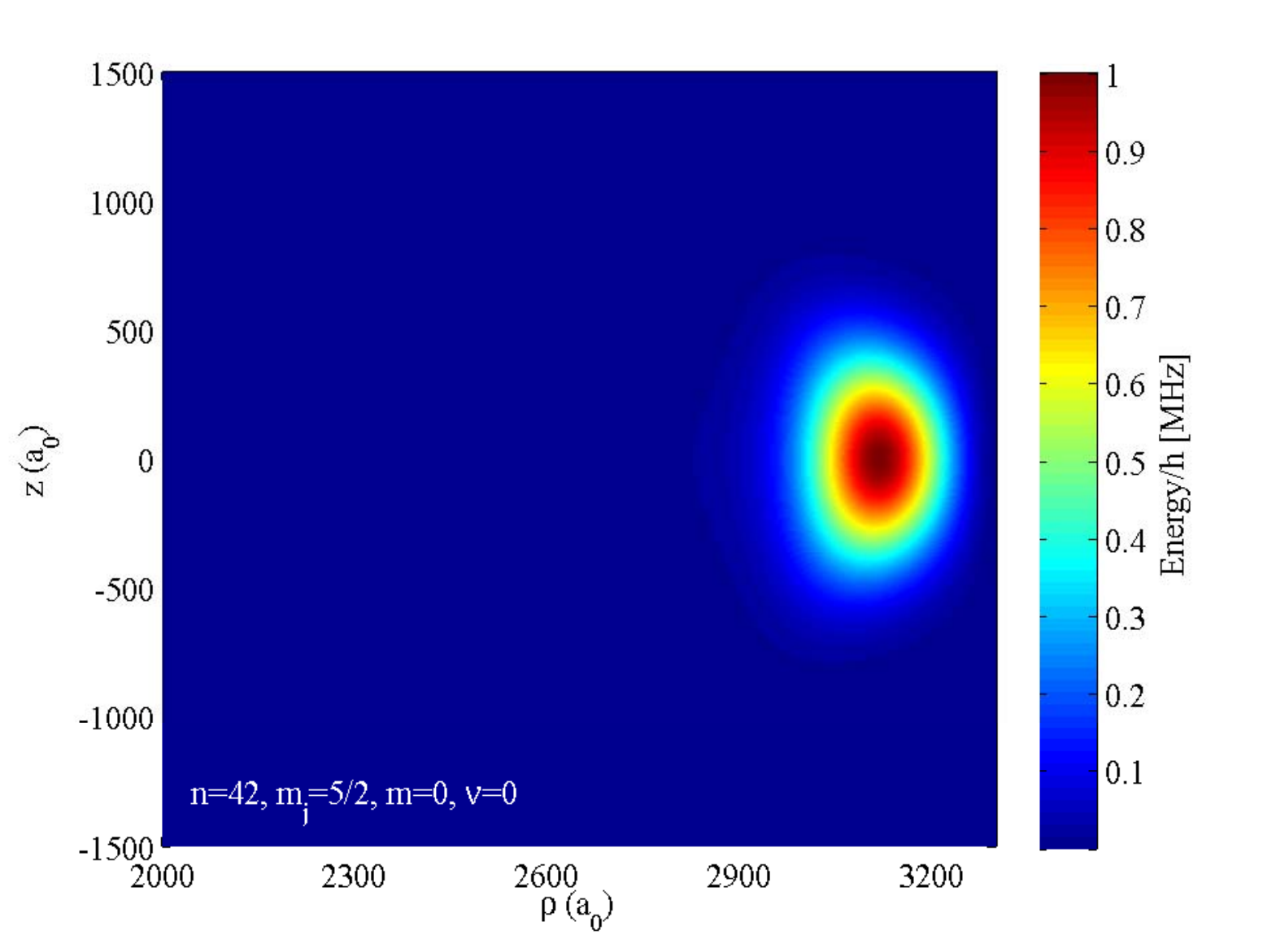}
\includegraphics[width=1.0\textwidth]{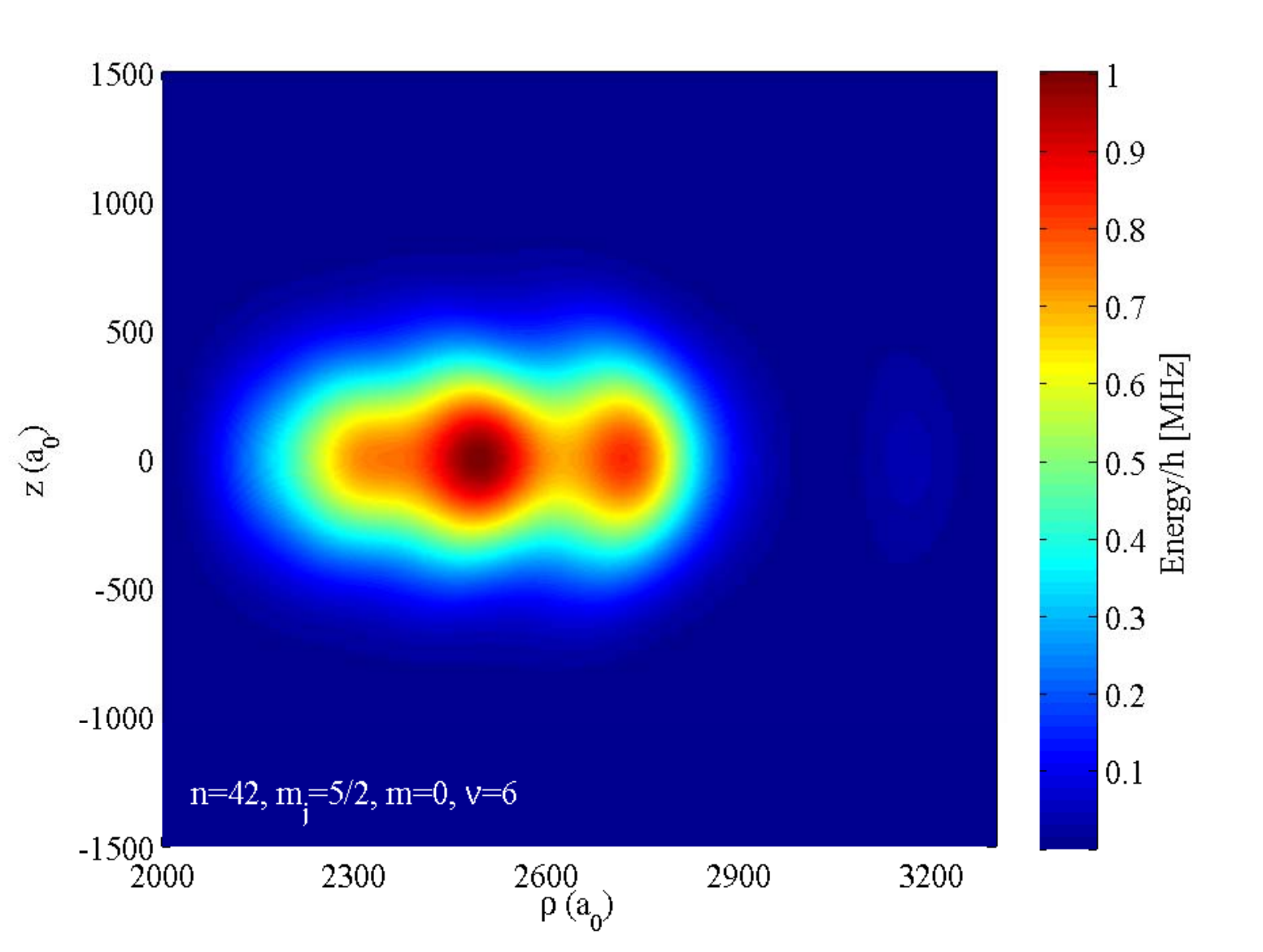}
\end{minipage}\qquad
\begin{minipage}[b]{.4\textwidth}
\includegraphics[width=1.0\textwidth]{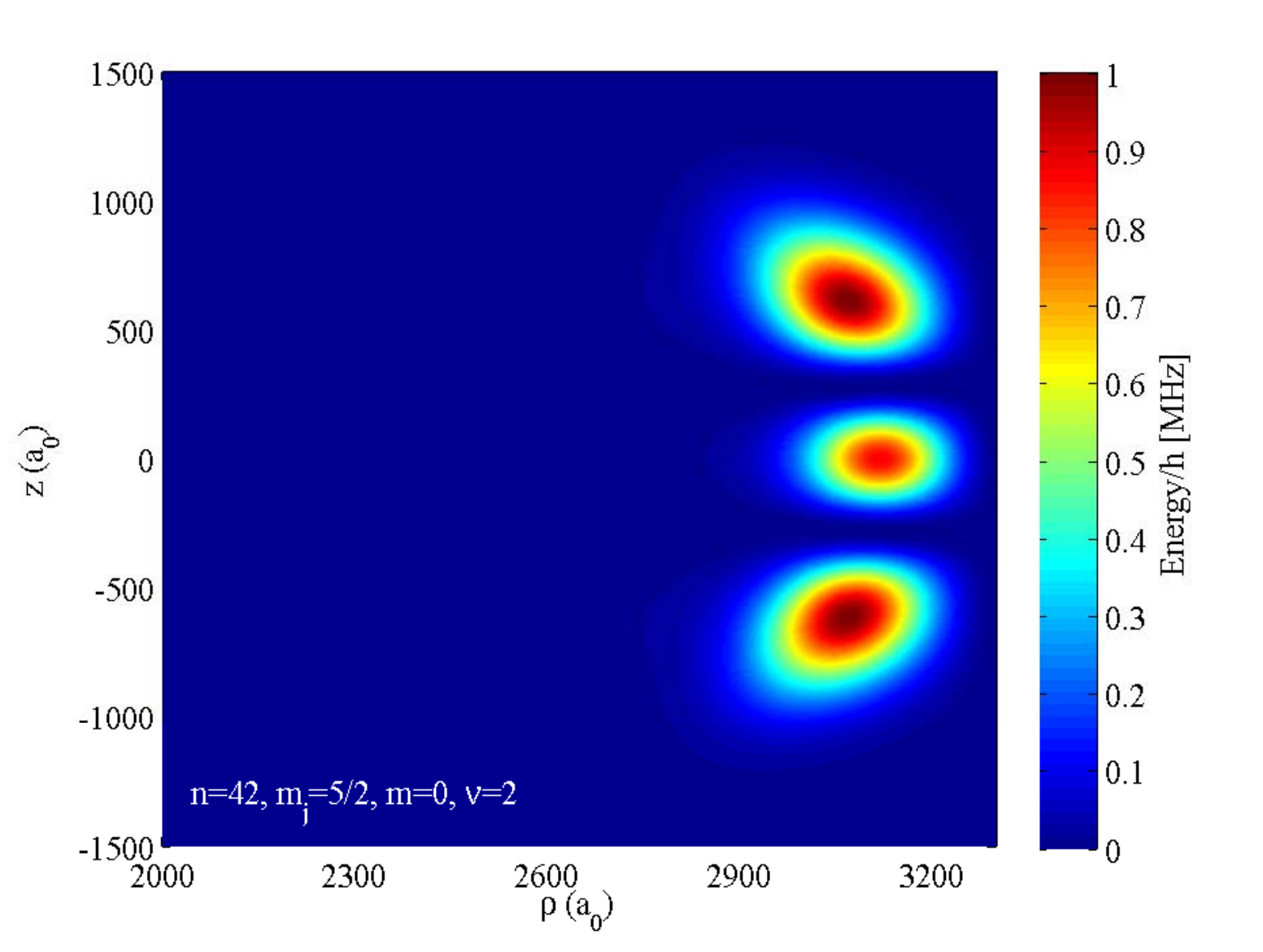}
\includegraphics[width=1.0\textwidth]{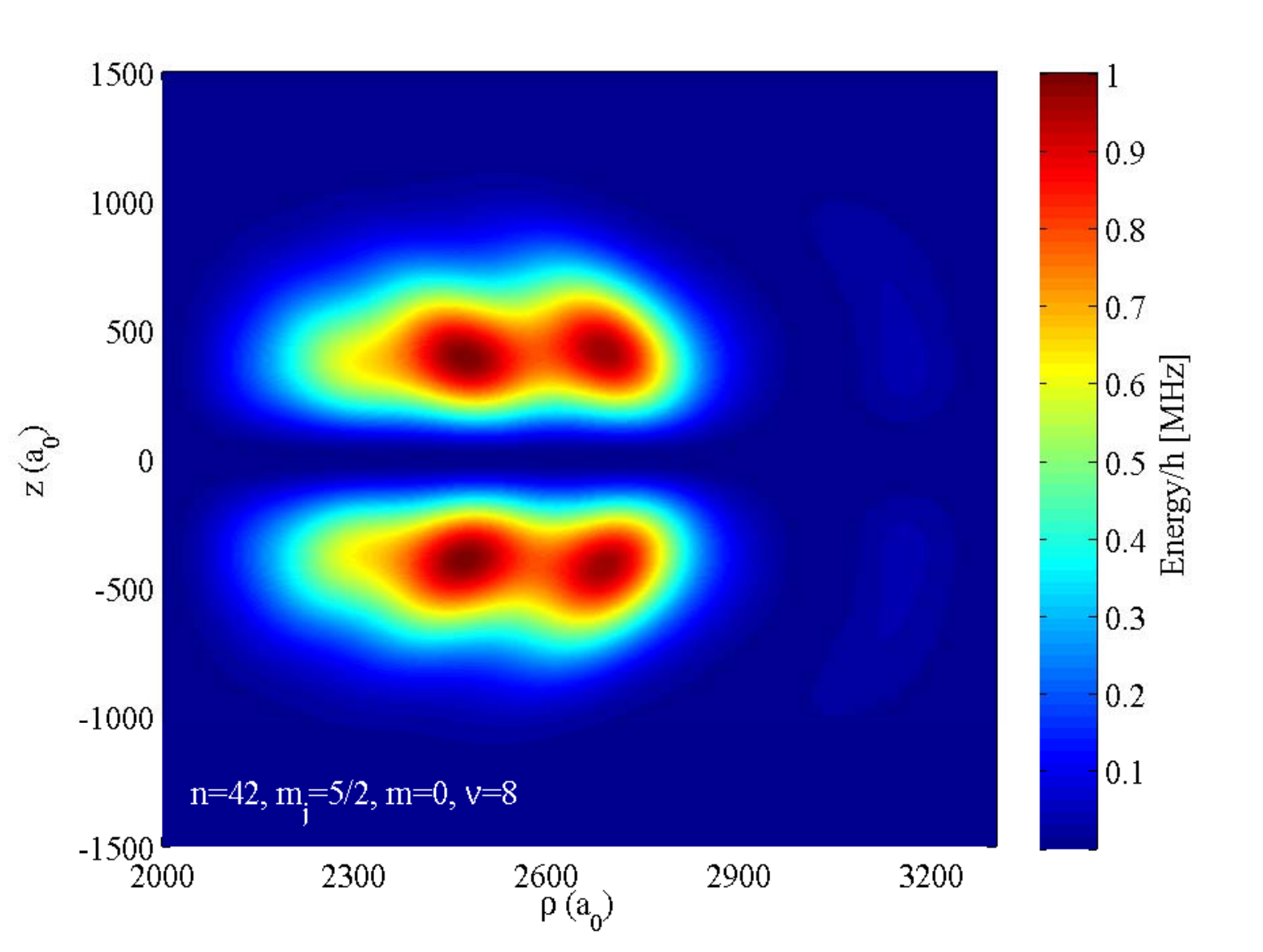}
\end{minipage}
\caption{(Scaled) rovibrational probability densities $|F_{\nu0}(\rho,z)|^2$ for $42D_{5/2},\ m_J=5/2$ APES.}\label{fig4}
\end{figure*}
\begin{figure*}[t]
\centering
\begin{minipage}[b]{.4\textwidth}
\includegraphics[width=1.0\textwidth]{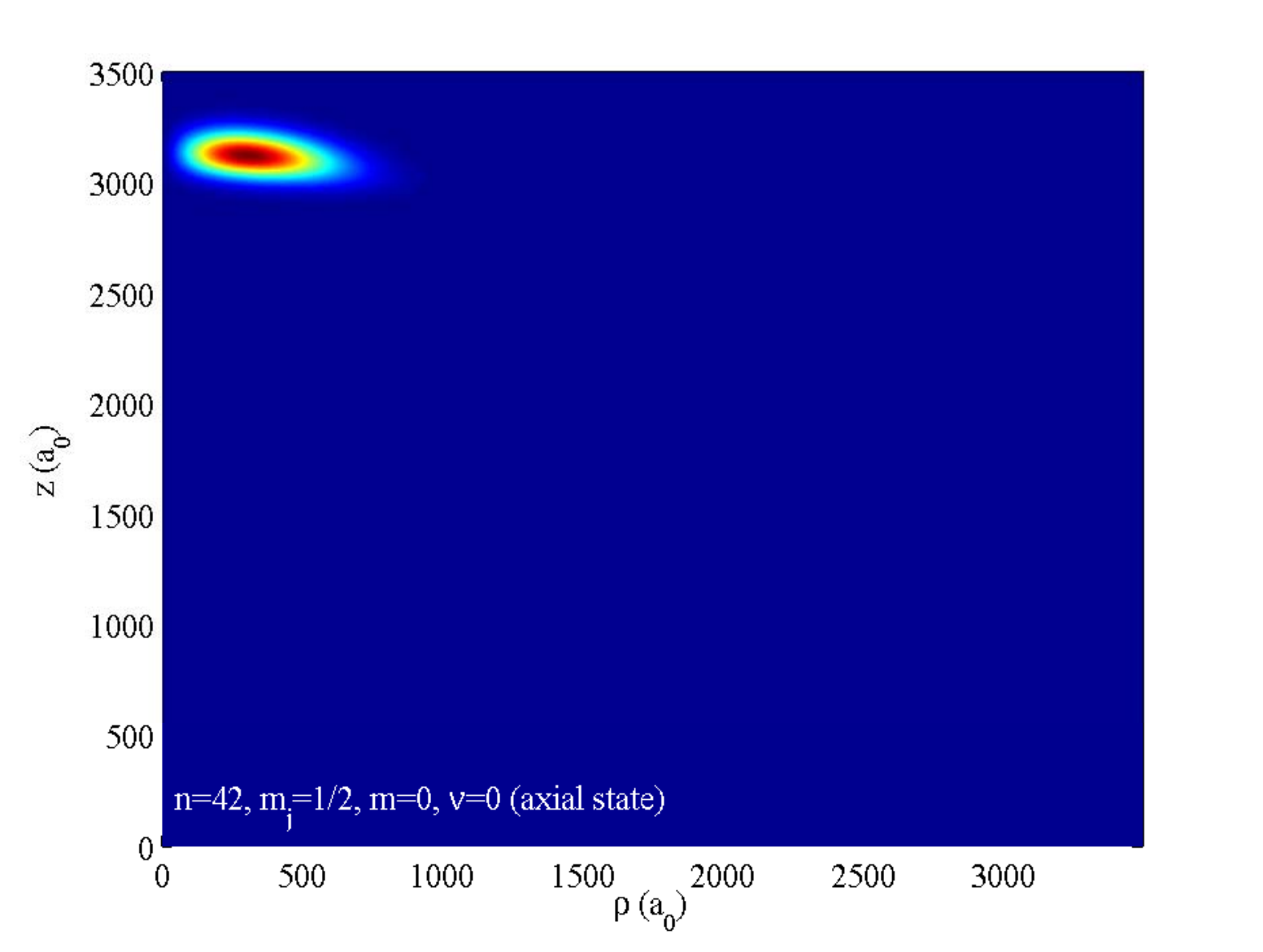}
\includegraphics[width=1.0\textwidth]{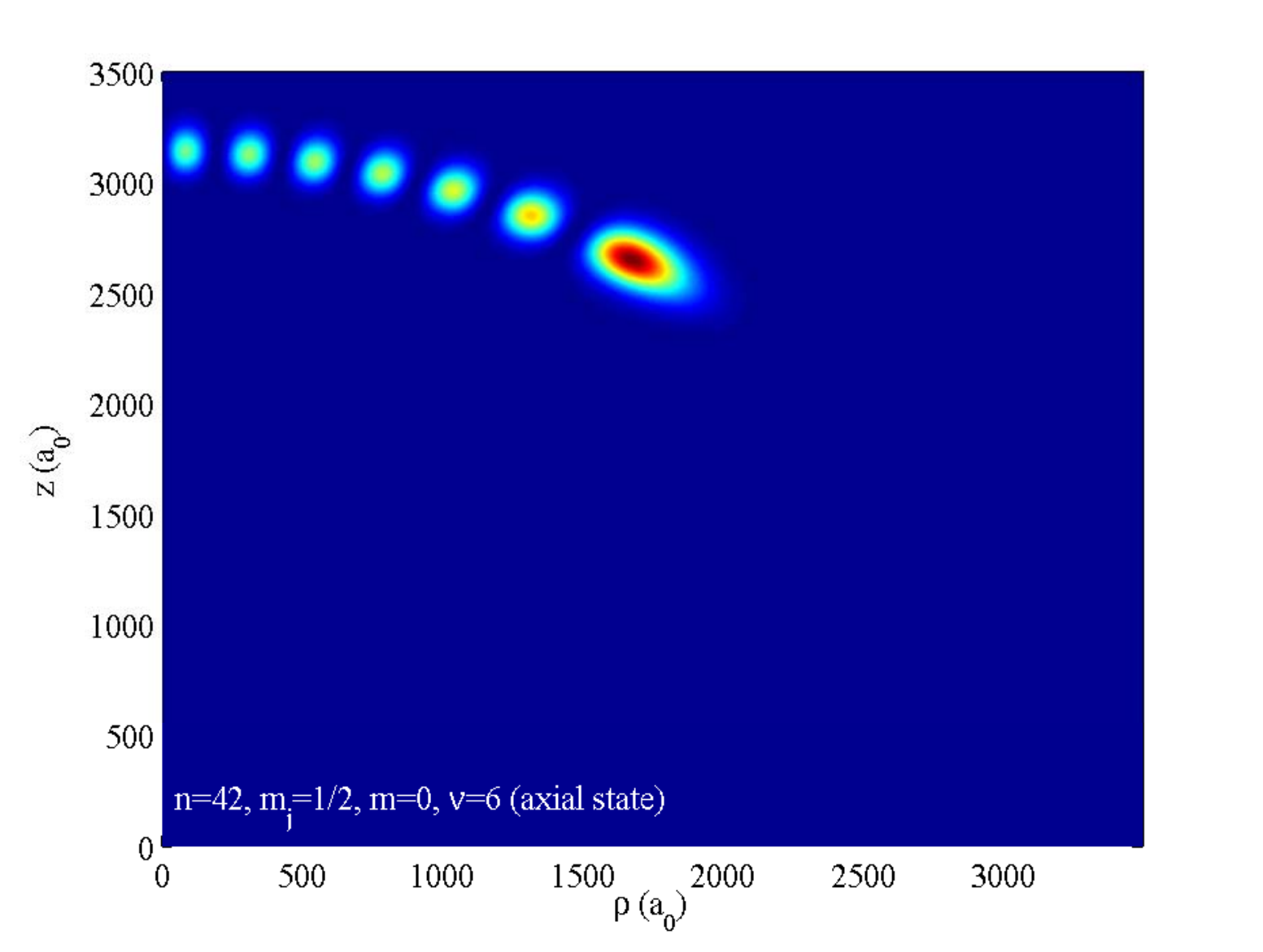}
\includegraphics[width=1.0\textwidth]{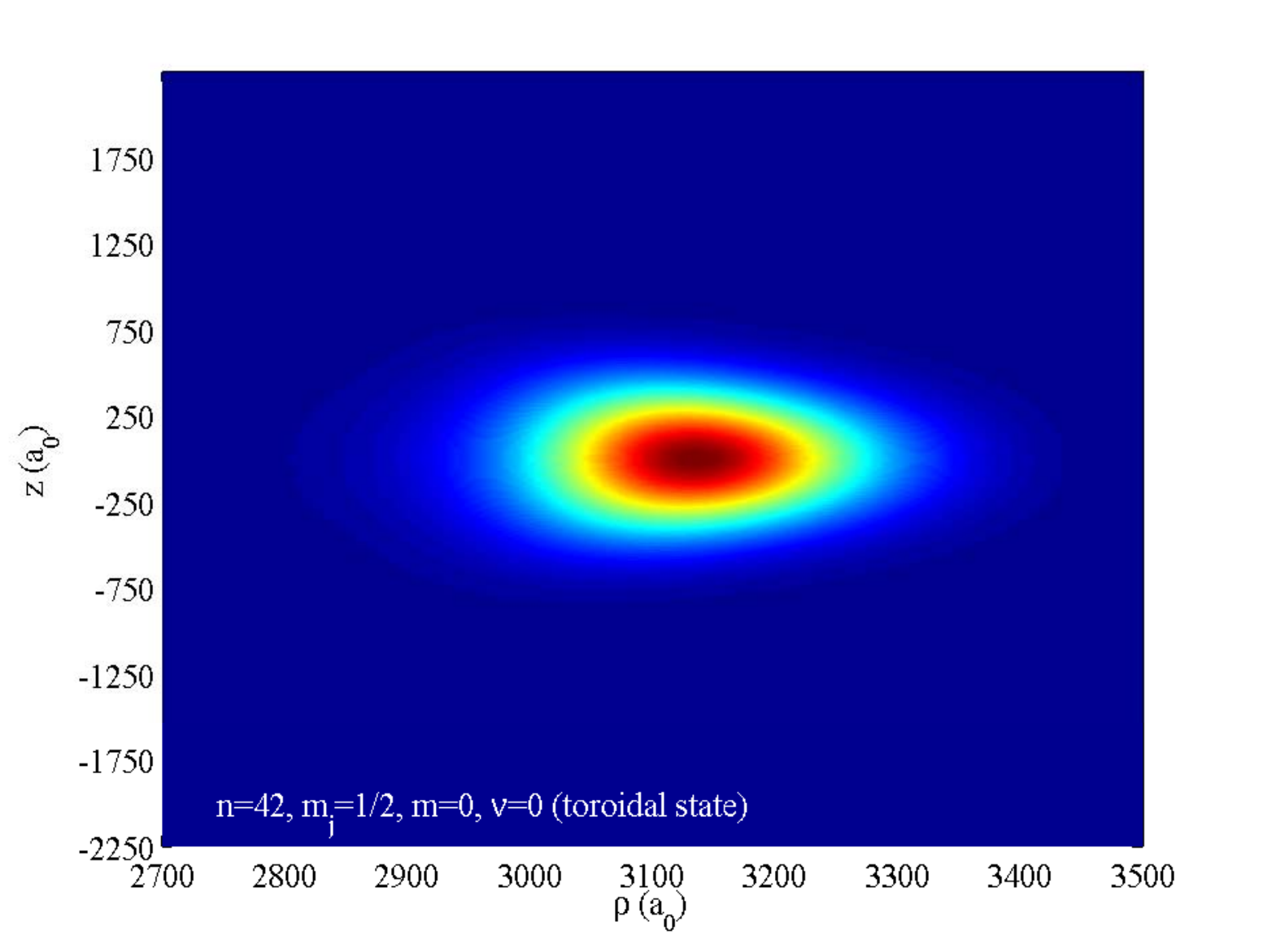}
\end{minipage}\qquad
\begin{minipage}[b]{.4\textwidth}
\includegraphics[width=1.0\textwidth]{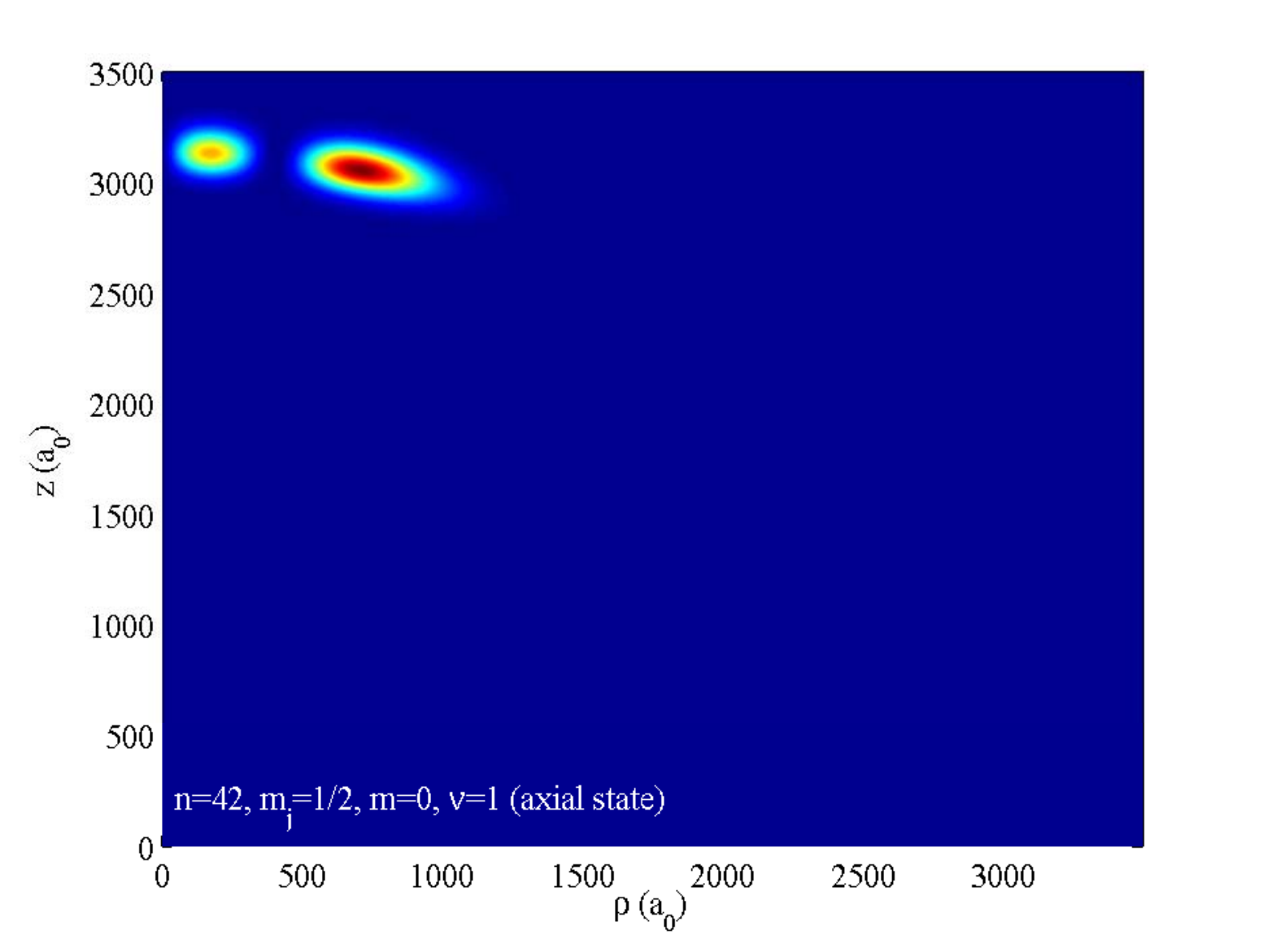}
\includegraphics[width=1.0\textwidth]{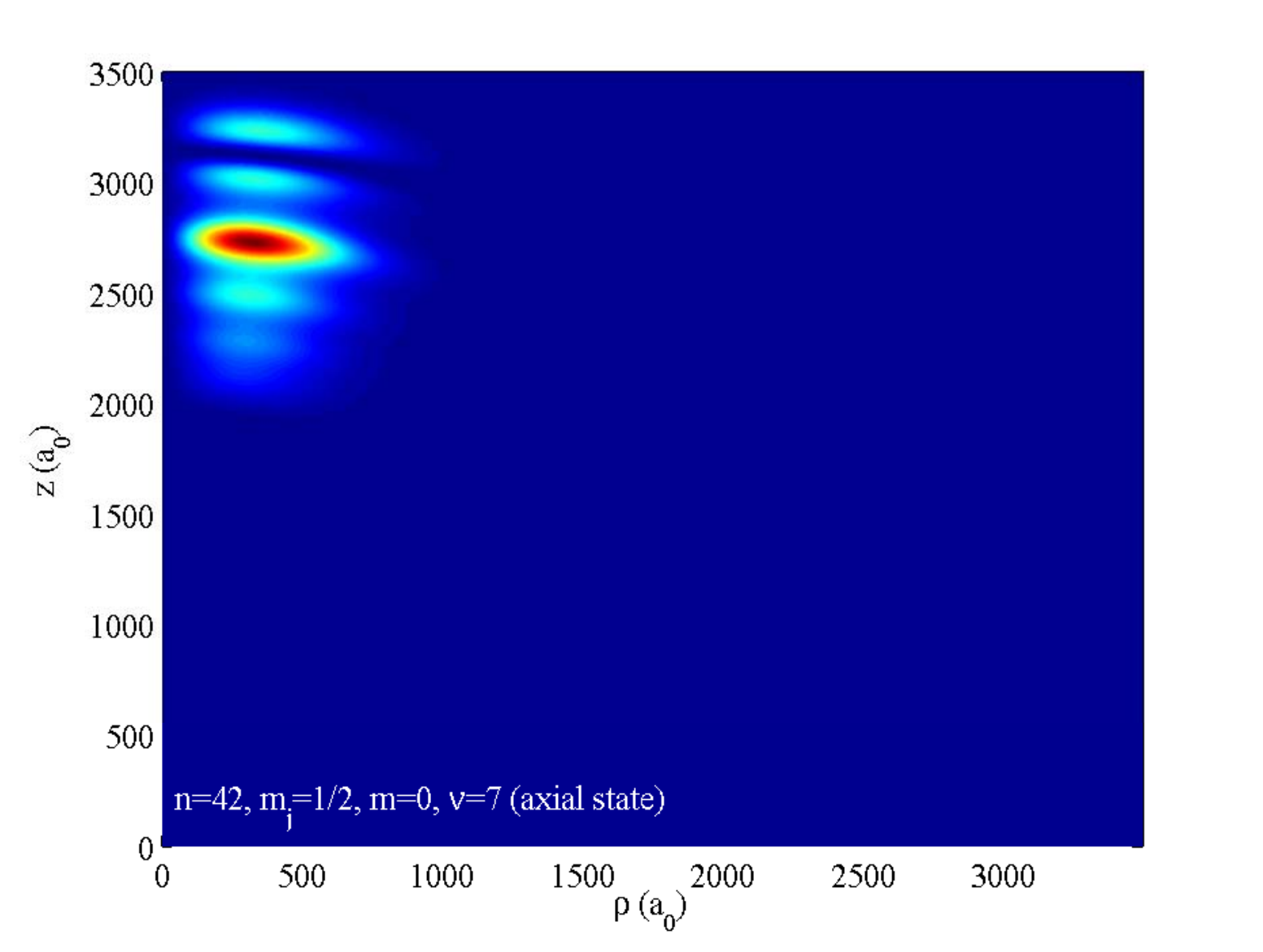}
\includegraphics[width=1.0\textwidth]{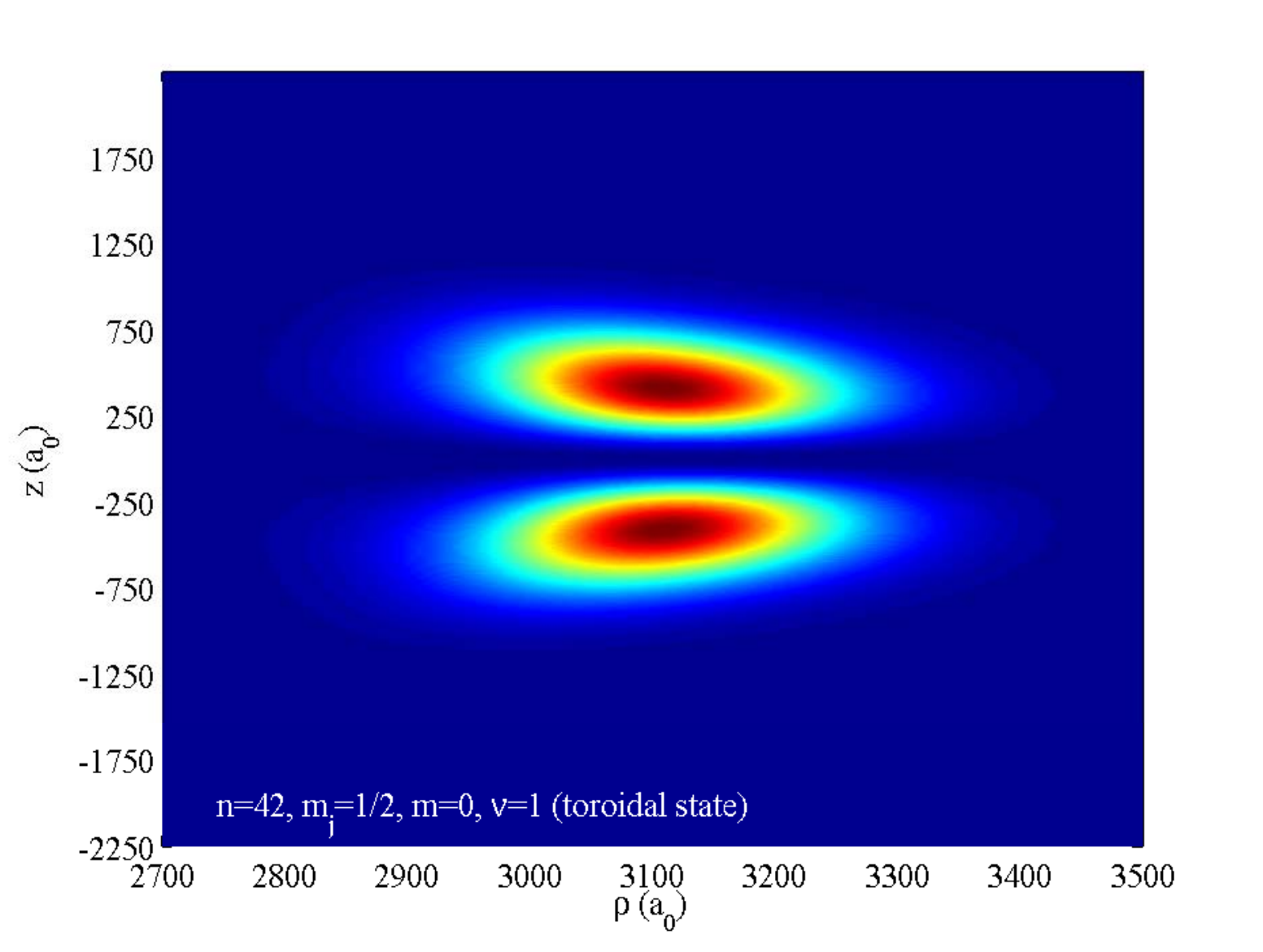}
\end{minipage}
\caption{(Scaled) probability densities $|F_{\nu0}(\rho,z)|^2$ for $42D_{5/2},\ m_J=1/2$ APES. We distinguish between axial ($\theta=0,\pi$) and toroidal states ($\theta=\pi/2$).}\label{fig5}
\end{figure*}

\begin{thebibliography}{1}
\makeatletter
\providecommand \@ifxundefined [1]{%
 \@ifx{#1\undefined}
}%
\providecommand \@ifnum [1]{%
 \ifnum #1\expandafter \@firstoftwo
 \else \expandafter \@secondoftwo
 \fi
}%
\providecommand \@ifx [1]{%
 \ifx #1\expandafter \@firstoftwo
 \else \expandafter \@secondoftwo
 \fi
}%
\providecommand \natexlab [1]{#1}%
\providecommand \enquote  [1]{``#1''}%
\providecommand \bibnamefont  [1]{#1}%
\providecommand \bibfnamefont [1]{#1}%
\providecommand \citenamefont [1]{#1}%
\providecommand \href@noop [0]{\@secondoftwo}%
\providecommand \href [0]{\begingroup \@sanitize@url \@href}%
\providecommand \@href[1]{\@@startlink{#1}\@@href}%
\providecommand \@@href[1]{\endgroup#1\@@endlink}%
\providecommand \@sanitize@url [0]{\catcode `\\12\catcode `\$12\catcode
  `\&12\catcode `\#12\catcode `\^12\catcode `\_12\catcode `\%12\relax}%
\providecommand \@@startlink[1]{}%
\providecommand \@@endlink[0]{}%
\providecommand \url  [0]{\begingroup\@sanitize@url \@url }%
\providecommand \@url [1]{\endgroup\@href {#1}{\urlprefix }}%
\providecommand \urlprefix  [0]{URL }%
\providecommand \Eprint [0]{\href }%
\providecommand \doibase [0]{http://dx.doi.org/}%
\providecommand \selectlanguage [0]{\@gobble}%
\providecommand \bibinfo  [0]{\@secondoftwo}%
\providecommand \bibfield  [0]{\@secondoftwo}%
\providecommand \translation [1]{[#1]}%
\providecommand \BibitemOpen [0]{}%
\providecommand \bibitemStop [0]{}%
\providecommand \bibitemNoStop [0]{.\EOS\space}%
\providecommand \EOS [0]{\spacefactor3000\relax}%
\providecommand \BibitemShut  [1]{\csname bibitem#1\endcsname}%
\let\auto@bib@innerbib\@empty
\bibitem [{\citenamefont {Brooks}\ and\ \citenamefont
  {Jones}(1966)}]{brooks:jcp45}%
  \BibitemOpen
  \bibfield  {author} {\bibinfo {author} {\bibfnamefont {P.~R.}\ \bibnamefont
  {Brooks}}\ and\ \bibinfo {author} {\bibfnamefont {E.~M.}\ \bibnamefont
  {Jones}},\ }\href@noop {} {\bibfield  {journal} {\bibinfo  {journal} {J.
  Chem. Phys.}\ }\textbf {\bibinfo {volume} {45}},\ \bibinfo {pages} {3449}
  (\bibinfo {year} {1966})}\BibitemShut {NoStop}%
\bibitem [{\citenamefont {Stolte}(1982)}]{Stolte:BBGPC86:413}%
  \BibitemOpen
  \bibfield  {author} {\bibinfo {author} {\bibfnamefont {S.}~\bibnamefont
  {Stolte}},\ }\href@noop {} {\bibfield  {journal} {\bibinfo  {journal} {Ber.
  Bunsen. Phys. Chem.}\ }\textbf {\bibinfo {volume} {86}},\ \bibinfo {pages}
  {413} (\bibinfo {year} {1982})}\BibitemShut {NoStop}%
\bibitem [{\citenamefont {Zare}(1998)}]{zare:science}%
  \BibitemOpen
  \bibfield  {author} {\bibinfo {author} {\bibfnamefont {R.}~\bibnamefont
  {Zare}},\ }\href@noop {} {\bibfield  {journal} {\bibinfo  {journal}
  {Science}\ }\textbf {\bibinfo {volume} {20}},\ \bibinfo {pages} {1875}
  (\bibinfo {year} {1998})}\BibitemShut {NoStop}%
\bibitem [{\citenamefont {Aquilanti}\ \emph {et~al.}(2005)\citenamefont
  {Aquilanti}, \citenamefont {Bartolomei}, \citenamefont {Pirani},
  \citenamefont {Cappelletti},\ and\ \citenamefont
  {Vecchiocattivi}}]{aquilanti:pccp_7}%
  \BibitemOpen
  \bibfield  {author} {\bibinfo {author} {\bibfnamefont {V.}~\bibnamefont
  {Aquilanti}}, \bibinfo {author} {\bibfnamefont {M.}~\bibnamefont
  {Bartolomei}}, \bibinfo {author} {\bibfnamefont {F.}~\bibnamefont {Pirani}},
  \bibinfo {author} {\bibfnamefont {D.}~\bibnamefont {Cappelletti}}, \ and\
  \bibinfo {author} {\bibfnamefont {F.}~\bibnamefont {Vecchiocattivi}},\
  }\href@noop {} {\bibfield  {journal} {\bibinfo  {journal} {Phys. Chem. Chem.
  Phys}\ }\textbf {\bibinfo {volume} {7}},\ \bibinfo {pages} {291} (\bibinfo
  {year} {2005})}\BibitemShut {NoStop}%
\bibitem [{\citenamefont {Holmegaard}\ \emph {et~al.}(2010)\citenamefont
  {Holmegaard}, \citenamefont {Hansen}, \citenamefont {Kalhoj}, \citenamefont
  {Kragh}, \citenamefont {Stapelfeldt}, \citenamefont {Filsinger},
  \citenamefont {K\"upper}, \citenamefont {Meijer}, \citenamefont
  {Dimitrovski}, \citenamefont {Abu-samha}, \citenamefont {Martiny},\ and\
  \citenamefont {Madsen}}]{Holmegaard:natphys6}%
  \BibitemOpen
  \bibfield  {author} {\bibinfo {author} {\bibfnamefont {L.}~\bibnamefont
  {Holmegaard}}, \bibinfo {author} {\bibfnamefont {J.~L.}\ \bibnamefont
  {Hansen}}, \bibinfo {author} {\bibfnamefont {L.}~\bibnamefont {Kalhoj}},
  \bibinfo {author} {\bibfnamefont {S.~L.}\ \bibnamefont {Kragh}}, \bibinfo
  {author} {\bibfnamefont {H.}~\bibnamefont {Stapelfeldt}}, \bibinfo {author}
  {\bibfnamefont {F.}~\bibnamefont {Filsinger}}, \bibinfo {author}
  {\bibfnamefont {J.}~\bibnamefont {K\"upper}}, \bibinfo {author}
  {\bibfnamefont {G.}~\bibnamefont {Meijer}}, \bibinfo {author} {\bibfnamefont
  {D.}~\bibnamefont {Dimitrovski}}, \bibinfo {author} {\bibfnamefont
  {M.}~\bibnamefont {Abu-samha}}, \bibinfo {author} {\bibfnamefont {C.~P.~J.}\
  \bibnamefont {Martiny}}, \ and\ \bibinfo {author} {\bibfnamefont {L.~B.}\
  \bibnamefont {Madsen}},\ }\href@noop {} {\bibfield  {journal} {\bibinfo
  {journal} {Nat. Phys.}\ }\textbf {\bibinfo {volume} {6}},\ \bibinfo {pages}
  {428} (\bibinfo {year} {2010})}\BibitemShut {NoStop}%
\bibitem [{\citenamefont {Hansen}\ \emph {et~al.}(2011)\citenamefont {Hansen},
  \citenamefont {Holmegaard}, \citenamefont {Kalh\o{}j}, \citenamefont {Kragh},
  \citenamefont {Stapelfeldt}, \citenamefont {Filsinger}, \citenamefont
  {Meijer}, \citenamefont {K\"upper}, \citenamefont {Dimitrovski},
  \citenamefont {Abu-samha}, \citenamefont {Martiny},\ and\ \citenamefont
  {Madsen}}]{PhysRevA.83.023406}%
  \BibitemOpen
  \bibfield  {author} {\bibinfo {author} {\bibfnamefont {J.~L.}\ \bibnamefont
  {Hansen}}, \bibinfo {author} {\bibfnamefont {L.}~\bibnamefont {Holmegaard}},
  \bibinfo {author} {\bibfnamefont {L.}~\bibnamefont {Kalh\o{}j}}, \bibinfo
  {author} {\bibfnamefont {S.~L.}\ \bibnamefont {Kragh}}, \bibinfo {author}
  {\bibfnamefont {H.}~\bibnamefont {Stapelfeldt}}, \bibinfo {author}
  {\bibfnamefont {F.}~\bibnamefont {Filsinger}}, \bibinfo {author}
  {\bibfnamefont {G.}~\bibnamefont {Meijer}}, \bibinfo {author} {\bibfnamefont
  {J.}~\bibnamefont {K\"upper}}, \bibinfo {author} {\bibfnamefont
  {D.}~\bibnamefont {Dimitrovski}}, \bibinfo {author} {\bibfnamefont
  {M.}~\bibnamefont {Abu-samha}}, \bibinfo {author} {\bibfnamefont {C.~P.~J.}\
  \bibnamefont {Martiny}}, \ and\ \bibinfo {author} {\bibfnamefont {L.~B.}\
  \bibnamefont {Madsen}},\ }\href@noop {} {\bibfield  {journal} {\bibinfo
  {journal} {Phys. Rev. A}\ }\textbf {\bibinfo {volume} {83}},\ \bibinfo
  {pages} {023406} (\bibinfo {year} {2011})}\BibitemShut {NoStop}%
\bibitem [{\citenamefont {Landers}\ \emph {et~al.}(2001)\citenamefont
  {Landers}, \citenamefont {Weber}, \citenamefont {Ali}, \citenamefont
  {Cassimi}, \citenamefont {Hattass}, \citenamefont {Jagutzki}, \citenamefont
  {Nauert}, \citenamefont {Osipov}, \citenamefont {Staudte}, \citenamefont
  {Prior}, \citenamefont {Schmidt-B{\"o}cking}, \citenamefont {Cocke},\ and\
  \citenamefont {D{\"o}rner}}]{Landers:PRL87:013002}%
  \BibitemOpen
  \bibfield  {author} {\bibinfo {author} {\bibfnamefont {A.}~\bibnamefont
  {Landers}}, \bibinfo {author} {\bibfnamefont {T.}~\bibnamefont {Weber}},
  \bibinfo {author} {\bibfnamefont {I.}~\bibnamefont {Ali}}, \bibinfo {author}
  {\bibfnamefont {A.}~\bibnamefont {Cassimi}}, \bibinfo {author} {\bibfnamefont
  {M.}~\bibnamefont {Hattass}}, \bibinfo {author} {\bibfnamefont
  {O.}~\bibnamefont {Jagutzki}}, \bibinfo {author} {\bibfnamefont
  {A.}~\bibnamefont {Nauert}}, \bibinfo {author} {\bibfnamefont
  {T.}~\bibnamefont {Osipov}}, \bibinfo {author} {\bibfnamefont
  {A.}~\bibnamefont {Staudte}}, \bibinfo {author} {\bibfnamefont
  {M.}~\bibnamefont {Prior}}, \bibinfo {author} {\bibfnamefont
  {H.}~\bibnamefont {Schmidt-B{\"o}cking}}, \bibinfo {author} {\bibfnamefont
  {C.}~\bibnamefont {Cocke}}, \ and\ \bibinfo {author} {\bibfnamefont
  {R.}~\bibnamefont {D{\"o}rner}},\ }\href@noop {} {\bibfield  {journal}
  {\bibinfo  {journal} {Phys. Rev. Lett.}\ }\textbf {\bibinfo {volume} {87}},\
  \bibinfo {pages} {013002} (\bibinfo {year} {2001})}\BibitemShut {NoStop}%
\bibitem [{\citenamefont {Bisgaard}\ \emph {et~al.}(2009)\citenamefont
  {Bisgaard}, \citenamefont {Clarkin}, \citenamefont {Wu}, \citenamefont {Lee},
  \citenamefont {Gessner}, \citenamefont {Hayden},\ and\ \citenamefont
  {Stolow}}]{bisgaard:science323}%
  \BibitemOpen
  \bibfield  {author} {\bibinfo {author} {\bibfnamefont {C.~Z.}\ \bibnamefont
  {Bisgaard}}, \bibinfo {author} {\bibfnamefont {O.~J.}\ \bibnamefont
  {Clarkin}}, \bibinfo {author} {\bibfnamefont {G.~R.}\ \bibnamefont {Wu}},
  \bibinfo {author} {\bibfnamefont {A.~M.~D.}\ \bibnamefont {Lee}}, \bibinfo
  {author} {\bibfnamefont {O.}~\bibnamefont {Gessner}}, \bibinfo {author}
  {\bibfnamefont {C.~C.}\ \bibnamefont {Hayden}}, \ and\ \bibinfo {author}
  {\bibfnamefont {A.}~\bibnamefont {Stolow}},\ }\href@noop {} {\bibfield
  {journal} {\bibinfo  {journal} {Science}\ }\textbf {\bibinfo {volume}
  {323}},\ \bibinfo {pages} {1464} (\bibinfo {year} {2009})}\BibitemShut
  {NoStop}%
\bibitem [{\citenamefont {Wu}\ \emph {et~al.}(1994)\citenamefont {Wu},
  \citenamefont {Bemish},\ and\ \citenamefont {Miller}}]{wu:jcp101}%
  \BibitemOpen
  \bibfield  {author} {\bibinfo {author} {\bibfnamefont {M.}~\bibnamefont
  {Wu}}, \bibinfo {author} {\bibfnamefont {R.~J.}\ \bibnamefont {Bemish}}, \
  and\ \bibinfo {author} {\bibfnamefont {R.~E.}\ \bibnamefont {Miller}},\
  }\href@noop {} {\bibfield  {journal} {\bibinfo  {journal} {J. Chem. Phys.}\
  }\textbf {\bibinfo {volume} {101}},\ \bibinfo {pages} {9447} (\bibinfo {year}
  {1994})}\BibitemShut {NoStop}%
\bibitem [{\citenamefont {Baumfalk}\ \emph {et~al.}(2001)\citenamefont
  {Baumfalk}, \citenamefont {Nahler},\ and\ \citenamefont
  {Buck}}]{baumfalk:jcp114}%
  \BibitemOpen
  \bibfield  {author} {\bibinfo {author} {\bibfnamefont {R.}~\bibnamefont
  {Baumfalk}}, \bibinfo {author} {\bibfnamefont {N.~H.}\ \bibnamefont
  {Nahler}}, \ and\ \bibinfo {author} {\bibfnamefont {U.}~\bibnamefont
  {Buck}},\ }\href@noop {} {\bibfield  {journal} {\bibinfo  {journal} {J. Chem.
  Phys.}\ }\textbf {\bibinfo {volume} {114}},\ \bibinfo {pages} {4755}
  (\bibinfo {year} {2001})}\BibitemShut {NoStop}%
\bibitem [{\citenamefont {van~den Brom}\ \emph {et~al.}(2004)\citenamefont
  {van~den Brom}, \citenamefont {Rakitzis},\ and\ \citenamefont
  {Janssen}}]{brom:11645}%
  \BibitemOpen
  \bibfield  {author} {\bibinfo {author} {\bibfnamefont {A.~J.}\ \bibnamefont
  {van~den Brom}}, \bibinfo {author} {\bibfnamefont {T.~P.}\ \bibnamefont
  {Rakitzis}}, \ and\ \bibinfo {author} {\bibfnamefont {M.~H.~M.}\ \bibnamefont
  {Janssen}},\ }\href@noop {} {\bibfield  {journal} {\bibinfo  {journal} {J.
  Chem. Phys.}\ }\textbf {\bibinfo {volume} {121}},\ \bibinfo {pages} {11645}
  (\bibinfo {year} {2004})}\BibitemShut {NoStop}%
\bibitem [{\citenamefont {Lipciuc}\ \emph {et~al.}(2005)\citenamefont
  {Lipciuc}, \citenamefont {van~den Brom}, \citenamefont {Dinu},\ and\
  \citenamefont {Janssen}}]{lipciuc:123103}%
  \BibitemOpen
  \bibfield  {author} {\bibinfo {author} {\bibfnamefont {M.~L.}\ \bibnamefont
  {Lipciuc}}, \bibinfo {author} {\bibfnamefont {A.~J.}\ \bibnamefont {van~den
  Brom}}, \bibinfo {author} {\bibfnamefont {L.}~\bibnamefont {Dinu}}, \ and\
  \bibinfo {author} {\bibfnamefont {M.~H.~M.}\ \bibnamefont {Janssen}},\
  }\href@noop {} {\bibfield  {journal} {\bibinfo  {journal} {Rev. Sci.
  Instrum.}\ }\textbf {\bibinfo {volume} {76}},\ \bibinfo {pages} {123103}
  (\bibinfo {year} {2005})}\BibitemShut {NoStop}%
\bibitem [{\citenamefont {Spence}\ and\ \citenamefont
  {Doak}(2004)}]{PhysRevLett.92.198102}%
  \BibitemOpen
  \bibfield  {author} {\bibinfo {author} {\bibfnamefont {J.~C.~H.}\
  \bibnamefont {Spence}}\ and\ \bibinfo {author} {\bibfnamefont {R.~B.}\
  \bibnamefont {Doak}},\ }\href@noop {} {\bibfield  {journal} {\bibinfo
  {journal} {Phys. Rev. Lett.}\ }\textbf {\bibinfo {volume} {92}},\ \bibinfo
  {pages} {198102} (\bibinfo {year} {2004})}\BibitemShut {NoStop}%
\bibitem [{\citenamefont {Filsinger}\ \emph {et~al.}(2011)\citenamefont
  {Filsinger}, \citenamefont {Meijer}, \citenamefont {Stapelfeldt},
  \citenamefont {Chapman},\ and\ \citenamefont
  {K\"upper}}]{Filsinger:PCCP13:2076}%
  \BibitemOpen
  \bibfield  {author} {\bibinfo {author} {\bibfnamefont {F.}~\bibnamefont
  {Filsinger}}, \bibinfo {author} {\bibfnamefont {G.}~\bibnamefont {Meijer}},
  \bibinfo {author} {\bibfnamefont {H.}~\bibnamefont {Stapelfeldt}}, \bibinfo
  {author} {\bibfnamefont {H.}~\bibnamefont {Chapman}}, \ and\ \bibinfo
  {author} {\bibfnamefont {J.}~\bibnamefont {K\"upper}},\ }\href@noop {}
  {\bibfield  {journal} {\bibinfo  {journal} {Phys. Chem. Chem. Phys.}\
  }\textbf {\bibinfo {volume} {13}},\ \bibinfo {pages} {2076} (\bibinfo {year}
  {2011})}\BibitemShut {NoStop}%
\bibitem [{\citenamefont {de~Miranda}\ \emph {et~al.}(2011)\citenamefont
  {de~Miranda}, \citenamefont {Chotia}, \citenamefont {Neyenhuis},
  \citenamefont {Wang}, \citenamefont {Qu{\'e}m{\'e}ner}, \citenamefont
  {Ospelkaus}, \citenamefont {Bohn}, \citenamefont {Ye},\ and\ \citenamefont
  {Jin}}]{Miranda:NatPhys}%
  \BibitemOpen
  \bibfield  {author} {\bibinfo {author} {\bibfnamefont {M.~H.~G.}\
  \bibnamefont {de~Miranda}}, \bibinfo {author} {\bibfnamefont
  {A.}~\bibnamefont {Chotia}}, \bibinfo {author} {\bibfnamefont
  {B.}~\bibnamefont {Neyenhuis}}, \bibinfo {author} {\bibfnamefont
  {D.}~\bibnamefont {Wang}}, \bibinfo {author} {\bibfnamefont {G.}~\bibnamefont
  {Qu{\'e}m{\'e}ner}}, \bibinfo {author} {\bibfnamefont {S.}~\bibnamefont
  {Ospelkaus}}, \bibinfo {author} {\bibfnamefont {J.~L.}\ \bibnamefont {Bohn}},
  \bibinfo {author} {\bibfnamefont {J.}~\bibnamefont {Ye}}, \ and\ \bibinfo
  {author} {\bibfnamefont {D.~S.}\ \bibnamefont {Jin}},\ }\href@noop {}
  {\bibfield  {journal} {\bibinfo  {journal} {Nature Physics}\ } (\bibinfo
  {year} {2011})}\BibitemShut {NoStop}%
\bibitem [{\citenamefont {Loesch}\ and\ \citenamefont
  {Remscheid}(1990)}]{Loesch1990}%
  \BibitemOpen
  \bibfield  {author} {\bibinfo {author} {\bibfnamefont {H.~J.}\ \bibnamefont
  {Loesch}}\ and\ \bibinfo {author} {\bibfnamefont {A.}~\bibnamefont
  {Remscheid}},\ }\href {\doibase http://dx.doi.org/10.1063/1.458668}
  {\bibfield  {journal} {\bibinfo  {journal} {J.Chem.Phys.}\ }\textbf {\bibinfo
  {volume} {93}},\ \bibinfo {pages} {4779} (\bibinfo {year}
  {1990})}\BibitemShut {NoStop}%
\bibitem [{\citenamefont {Brooks}(1976)}]{brooks:science}%
  \BibitemOpen
  \bibfield  {author} {\bibinfo {author} {\bibfnamefont {P.~R.}\ \bibnamefont
  {Brooks}},\ }\href@noop {} {\bibfield  {journal} {\bibinfo  {journal}
  {Science}\ }\textbf {\bibinfo {volume} {193}},\ \bibinfo {pages} {11}
  (\bibinfo {year} {1976})}\BibitemShut {NoStop}%
\bibitem [{\citenamefont {Parker}\ and\ \citenamefont
  {Bernstein}(1989)}]{parker:1989}%
  \BibitemOpen
  \bibfield  {author} {\bibinfo {author} {\bibfnamefont {D.~H.}\ \bibnamefont
  {Parker}}\ and\ \bibinfo {author} {\bibfnamefont {R.~B.}\ \bibnamefont
  {Bernstein}},\ }\href@noop {} {\bibfield  {journal} {\bibinfo  {journal}
  {Annu. Rev. Phys. Chem.}\ }\textbf {\bibinfo {volume} {40}},\ \bibinfo
  {pages} {561} (\bibinfo {year} {1989})}\BibitemShut {NoStop}%
\bibitem [{\citenamefont {Hain}\ \emph {et~al.}(1999)\citenamefont {Hain},
  \citenamefont {Moision},\ and\ \citenamefont {Curtiss}}]{Hain1999}%
  \BibitemOpen
  \bibfield  {author} {\bibinfo {author} {\bibfnamefont {T.~D.}\ \bibnamefont
  {Hain}}, \bibinfo {author} {\bibfnamefont {R.~M.}\ \bibnamefont {Moision}}, \
  and\ \bibinfo {author} {\bibfnamefont {T.~J.}\ \bibnamefont {Curtiss}},\
  }\href {\doibase http://dx.doi.org/10.1063/1.480043} {\bibfield  {journal}
  {\bibinfo  {journal} {J.Chem.Phys.}\ }\textbf {\bibinfo {volume} {111}},\
  \bibinfo {pages} {6797} (\bibinfo {year} {1999})}\BibitemShut {NoStop}%
\bibitem [{\citenamefont {Stapelfeldt}\ and\ \citenamefont
  {Seideman}(2003)}]{Seide2003}%
  \BibitemOpen
  \bibfield  {author} {\bibinfo {author} {\bibfnamefont {H.}~\bibnamefont
  {Stapelfeldt}}\ and\ \bibinfo {author} {\bibfnamefont {T.}~\bibnamefont
  {Seideman}},\ }\href {\doibase 10.1103/RevModPhys.75.543} {\bibfield
  {journal} {\bibinfo  {journal} {Rev. Mod. Phys.}\ }\textbf {\bibinfo {volume}
  {75}},\ \bibinfo {pages} {543} (\bibinfo {year} {2003})}\BibitemShut
  {NoStop}%
\bibitem [{\citenamefont {Friedrich}\ and\ \citenamefont
  {Herschbach}(1999{\natexlab{a}})}]{friedrich:jcp111}%
  \BibitemOpen
  \bibfield  {author} {\bibinfo {author} {\bibfnamefont {B.}~\bibnamefont
  {Friedrich}}\ and\ \bibinfo {author} {\bibfnamefont {D.~R.}\ \bibnamefont
  {Herschbach}},\ }\href@noop {} {\bibfield  {journal} {\bibinfo  {journal} {J.
  Chem. Phys.}\ }\textbf {\bibinfo {volume} {111}},\ \bibinfo {pages} {6157}
  (\bibinfo {year} {1999}{\natexlab{a}})}\BibitemShut {NoStop}%
\bibitem [{\citenamefont {Friedrich}\ and\ \citenamefont
  {Herschbach}(1999{\natexlab{b}})}]{friedrich:jpca103}%
  \BibitemOpen
  \bibfield  {author} {\bibinfo {author} {\bibfnamefont {B.}~\bibnamefont
  {Friedrich}}\ and\ \bibinfo {author} {\bibfnamefont {D.~R.}\ \bibnamefont
  {Herschbach}},\ }\href@noop {} {\bibfield  {journal} {\bibinfo  {journal} {J.
  Phys. Chem. A}\ }\textbf {\bibinfo {volume} {103}},\ \bibinfo {pages} {10280}
  (\bibinfo {year} {1999}{\natexlab{b}})}\BibitemShut {NoStop}%
\bibitem [{\citenamefont {Sakai}\ \emph {et~al.}(2003)\citenamefont {Sakai},
  \citenamefont {Minemoto}, \citenamefont {Nanjo}, \citenamefont {Tanji},\ and\
  \citenamefont {Suzuki}}]{sakai:prl_90}%
  \BibitemOpen
  \bibfield  {author} {\bibinfo {author} {\bibfnamefont {H.}~\bibnamefont
  {Sakai}}, \bibinfo {author} {\bibfnamefont {S.}~\bibnamefont {Minemoto}},
  \bibinfo {author} {\bibfnamefont {H.}~\bibnamefont {Nanjo}}, \bibinfo
  {author} {\bibfnamefont {H.}~\bibnamefont {Tanji}}, \ and\ \bibinfo {author}
  {\bibfnamefont {T.}~\bibnamefont {Suzuki}},\ }\href@noop {} {\bibfield
  {journal} {\bibinfo  {journal} {Phys. Rev. Lett.}\ }\textbf {\bibinfo
  {volume} {90}},\ \bibinfo {pages} {083001} (\bibinfo {year}
  {2003})}\BibitemShut {NoStop}%
\bibitem [{\citenamefont {Tanji}\ \emph {et~al.}(2005)\citenamefont {Tanji},
  \citenamefont {Minemoto},\ and\ \citenamefont {Sakai}}]{Tanji2005}%
  \BibitemOpen
  \bibfield  {author} {\bibinfo {author} {\bibfnamefont {H.}~\bibnamefont
  {Tanji}}, \bibinfo {author} {\bibfnamefont {S.}~\bibnamefont {Minemoto}}, \
  and\ \bibinfo {author} {\bibfnamefont {H.}~\bibnamefont {Sakai}},\
  }\href@noop {} {\bibfield  {journal} {\bibinfo  {journal} {Phys. Rev. A}\
  }\textbf {\bibinfo {volume} {72}},\ \bibinfo {pages} {063401} (\bibinfo
  {year} {2005})}\BibitemShut {NoStop}%
\bibitem [{Kap()}]{Kappes}%
  \BibitemOpen
  \href@noop {} {}\bibinfo {note} {U.~Kappes and P.~Schmelcher, Phys. Lett. A
  \textbf{210}, 409 (1996); Phys. Rev. A \textbf{53}, 3869 (1996)}\BibitemShut
  {NoStop}%
\bibitem [{Det()}]{Detmer}%
  \BibitemOpen
  \href@noop {} {}\bibinfo {note} {T.~Detmer, P.~Schmelcher and L.S.~Cederbaum,
  Phys. Rev. A \textbf{57}, 1767 (1998); {\it{Atoms and Molecules in Strong
  External Fields}}, Editors: P.~Schmelcher and W.~Schweizer, {\bf{Plenum
  Publishing Company}} 1998.}\BibitemShut {Stop}%
\bibitem [{\citenamefont {Lange}\ \emph {et~al.}(2012)\citenamefont {Lange},
  \citenamefont {Tellgren}, \citenamefont {Hoffmann},\ and\ \citenamefont
  {Helgaker}}]{Helgaker2012}%
  \BibitemOpen
  \bibfield  {author} {\bibinfo {author} {\bibfnamefont {K.~K.}\ \bibnamefont
  {Lange}}, \bibinfo {author} {\bibfnamefont {E.~I.}\ \bibnamefont {Tellgren}},
  \bibinfo {author} {\bibfnamefont {M.~R.}\ \bibnamefont {Hoffmann}}, \ and\
  \bibinfo {author} {\bibfnamefont {T.}~\bibnamefont {Helgaker}},\ }\href
  {\doibase 10.1126/science.1219703} {\bibfield  {journal} {\bibinfo  {journal}
  {Science}\ }\textbf {\bibinfo {volume} {337}},\ \bibinfo {pages} {327}
  (\bibinfo {year} {2012})}\BibitemShut {NoStop}%
\bibitem [{\citenamefont {Schmelcher}\ and\ \citenamefont
  {Cederbaum}(1990)}]{Schmelcher1990}%
  \BibitemOpen
  \bibfield  {author} {\bibinfo {author} {\bibfnamefont {P.}~\bibnamefont
  {Schmelcher}}\ and\ \bibinfo {author} {\bibfnamefont {L.~S.}\ \bibnamefont
  {Cederbaum}},\ }\href {\doibase 10.1103/PhysRevA.41.4936} {\bibfield
  {journal} {\bibinfo  {journal} {Phys. Rev. A}\ }\textbf {\bibinfo {volume}
  {41}},\ \bibinfo {pages} {4936} (\bibinfo {year} {1990})}\BibitemShut
  {NoStop}%
\bibitem [{\citenamefont {Greene}\ \emph {et~al.}(2000)\citenamefont {Greene},
  \citenamefont {Dickinson},\ and\ \citenamefont {Sadeghpour}}]{Greene2000}%
  \BibitemOpen
  \bibfield  {author} {\bibinfo {author} {\bibfnamefont {C.~H.}\ \bibnamefont
  {Greene}}, \bibinfo {author} {\bibfnamefont {A.~S.}\ \bibnamefont
  {Dickinson}}, \ and\ \bibinfo {author} {\bibfnamefont {H.~R.}\ \bibnamefont
  {Sadeghpour}},\ }\href {\doibase 10.1103/PhysRevLett.85.2458} {\bibfield
  {journal} {\bibinfo  {journal} {Phys. Rev. Lett.}\ }\textbf {\bibinfo
  {volume} {85}},\ \bibinfo {pages} {2458} (\bibinfo {year}
  {2000})}\BibitemShut {NoStop}%
\bibitem [{\citenamefont {Lesanovsky}\ \emph {et~al.}(2006)\citenamefont
  {Lesanovsky}, \citenamefont {Schmelcher},\ and\ \citenamefont
  {Sadeghpour}}]{Lesanovsky2006}%
  \BibitemOpen
  \bibfield  {author} {\bibinfo {author} {\bibfnamefont {I.}~\bibnamefont
  {Lesanovsky}}, \bibinfo {author} {\bibfnamefont {P.}~\bibnamefont
  {Schmelcher}}, \ and\ \bibinfo {author} {\bibfnamefont {H.~R.}\ \bibnamefont
  {Sadeghpour}},\ }\href {http://stacks.iop.org/0953-4075/39/i=4/a=L03}
  {\bibfield  {journal} {\bibinfo  {journal} {Journal of Physics B: Atomic,
  Molecular and Optical Physics}\ }\textbf {\bibinfo {volume} {39}},\ \bibinfo
  {pages} {L69} (\bibinfo {year} {2006})}\BibitemShut {NoStop}%
\bibitem [{\citenamefont {Bendkowsky}\ \emph {et~al.}(2009)\citenamefont
  {Bendkowsky}, \citenamefont {Butscher}, \citenamefont {Nipper}, \citenamefont
  {Shaffer}, \citenamefont {L\"ow},\ and\ \citenamefont
  {Pfau}}]{Bendkowsky2009}%
  \BibitemOpen
  \bibfield  {author} {\bibinfo {author} {\bibfnamefont {V.}~\bibnamefont
  {Bendkowsky}}, \bibinfo {author} {\bibfnamefont {B.}~\bibnamefont
  {Butscher}}, \bibinfo {author} {\bibfnamefont {J.}~\bibnamefont {Nipper}},
  \bibinfo {author} {\bibfnamefont {J.~P.}\ \bibnamefont {Shaffer}}, \bibinfo
  {author} {\bibfnamefont {R.}~\bibnamefont {L\"ow}}, \ and\ \bibinfo {author}
  {\bibfnamefont {T.}~\bibnamefont {Pfau}},\ }\href
  {http://dx.doi.org/10.1038/nature07945} {\bibfield  {journal} {\bibinfo
  {journal} {Nature}\ }\textbf {\bibinfo {volume} {458}},\ \bibinfo {pages}
  {0028} (\bibinfo {year} {2009})}\BibitemShut {NoStop}%
\bibitem [{\citenamefont {Bendkowsky}\ \emph {et~al.}(2010)\citenamefont
  {Bendkowsky}, \citenamefont {Butscher}, \citenamefont {Nipper}, \citenamefont
  {Balewski}, \citenamefont {Shaffer}, \citenamefont {L\"ow}, \citenamefont
  {Pfau}, \citenamefont {Li}, \citenamefont {Stanojevic}, \citenamefont
  {Pohl},\ and\ \citenamefont {Rost}}]{Bendkowsky2010}%
  \BibitemOpen
  \bibfield  {author} {\bibinfo {author} {\bibfnamefont {V.}~\bibnamefont
  {Bendkowsky}}, \bibinfo {author} {\bibfnamefont {B.}~\bibnamefont
  {Butscher}}, \bibinfo {author} {\bibfnamefont {J.}~\bibnamefont {Nipper}},
  \bibinfo {author} {\bibfnamefont {J.~B.}\ \bibnamefont {Balewski}}, \bibinfo
  {author} {\bibfnamefont {J.~P.}\ \bibnamefont {Shaffer}}, \bibinfo {author}
  {\bibfnamefont {R.}~\bibnamefont {L\"ow}}, \bibinfo {author} {\bibfnamefont
  {T.}~\bibnamefont {Pfau}}, \bibinfo {author} {\bibfnamefont {W.}~\bibnamefont
  {Li}}, \bibinfo {author} {\bibfnamefont {J.}~\bibnamefont {Stanojevic}},
  \bibinfo {author} {\bibfnamefont {T.}~\bibnamefont {Pohl}}, \ and\ \bibinfo
  {author} {\bibfnamefont {J.~M.}\ \bibnamefont {Rost}},\ }\href {\doibase
  10.1103/PhysRevLett.105.163201} {\bibfield  {journal} {\bibinfo  {journal}
  {Phys. Rev. Lett.}\ }\textbf {\bibinfo {volume} {105}},\ \bibinfo {pages}
  {163201} (\bibinfo {year} {2010})}\BibitemShut {NoStop}%
\bibitem [{\citenamefont {Tallant}\ \emph {et~al.}(2012)\citenamefont
  {Tallant}, \citenamefont {Rittenhouse}, \citenamefont {Booth}, \citenamefont
  {Sadeghpour},\ and\ \citenamefont {Shaffer}}]{Tallant2012}%
  \BibitemOpen
  \bibfield  {author} {\bibinfo {author} {\bibfnamefont {J.}~\bibnamefont
  {Tallant}}, \bibinfo {author} {\bibfnamefont {S.~T.}\ \bibnamefont
  {Rittenhouse}}, \bibinfo {author} {\bibfnamefont {D.}~\bibnamefont {Booth}},
  \bibinfo {author} {\bibfnamefont {H.~R.}\ \bibnamefont {Sadeghpour}}, \ and\
  \bibinfo {author} {\bibfnamefont {J.~P.}\ \bibnamefont {Shaffer}},\ }\href
  {\doibase 10.1103/PhysRevLett.109.173202} {\bibfield  {journal} {\bibinfo
  {journal} {Phys. Rev. Lett.}\ }\textbf {\bibinfo {volume} {109}},\ \bibinfo
  {pages} {173202} (\bibinfo {year} {2012})}\BibitemShut {NoStop}%
\bibitem [{\citenamefont {Bellos}\ \emph {et~al.}(2013)\citenamefont {Bellos},
  \citenamefont {Carollo}, \citenamefont {Banerjee}, \citenamefont {Eyler},
  \citenamefont {Gould},\ and\ \citenamefont {Stwalley}}]{Bellos2013}%
  \BibitemOpen
  \bibfield  {author} {\bibinfo {author} {\bibfnamefont {M.~A.}\ \bibnamefont
  {Bellos}}, \bibinfo {author} {\bibfnamefont {R.}~\bibnamefont {Carollo}},
  \bibinfo {author} {\bibfnamefont {J.}~\bibnamefont {Banerjee}}, \bibinfo
  {author} {\bibfnamefont {E.~E.}\ \bibnamefont {Eyler}}, \bibinfo {author}
  {\bibfnamefont {P.~L.}\ \bibnamefont {Gould}}, \ and\ \bibinfo {author}
  {\bibfnamefont {W.~C.}\ \bibnamefont {Stwalley}},\ }\href {\doibase
  10.1103/PhysRevLett.111.053001} {\bibfield  {journal} {\bibinfo  {journal}
  {Phys. Rev. Lett.}\ }\textbf {\bibinfo {volume} {111}},\ \bibinfo {pages}
  {053001} (\bibinfo {year} {2013})}\BibitemShut {NoStop}%
\bibitem [{\citenamefont {Fermi}(1934)}]{Fermi1934}%
  \BibitemOpen
  \bibfield  {author} {\bibinfo {author} {\bibfnamefont {E.}~\bibnamefont
  {Fermi}},\ }\href@noop {} {\bibfield  {journal} {\bibinfo  {journal} {Nuovo
  Cimento}\ }\textbf {\bibinfo {volume} {11}},\ \bibinfo {pages} {157}
  (\bibinfo {year} {1934})}\BibitemShut {NoStop}%
\bibitem [{\citenamefont {Omont}(1977)}]{Omont1977}%
  \BibitemOpen
  \bibfield  {author} {\bibinfo {author} {\bibfnamefont {A.}~\bibnamefont
  {Omont}},\ }\href {\doibase 10.1051/jphys:0197700380110134300} {\bibfield
  {journal} {\bibinfo  {journal} {J. Phys. France}\ }\textbf {\bibinfo {volume}
  {38}},\ \bibinfo {pages} {1343} (\bibinfo {year} {1977})}\BibitemShut
  {NoStop}%
\bibitem [{\citenamefont {L\"ow}\ \emph {et~al.}(2012)\citenamefont {L\"ow},
  \citenamefont {Weimer}, \citenamefont {Nipper}, \citenamefont {Balewski},
  \citenamefont {Butscher}, \citenamefont {B\"uchler},\ and\ \citenamefont
  {Pfau}}]{LWN12}%
  \BibitemOpen
  \bibfield  {author} {\bibinfo {author} {\bibfnamefont {R.}~\bibnamefont
  {L\"ow}}, \bibinfo {author} {\bibfnamefont {H.}~\bibnamefont {Weimer}},
  \bibinfo {author} {\bibfnamefont {J.}~\bibnamefont {Nipper}}, \bibinfo
  {author} {\bibfnamefont {J.~B.}\ \bibnamefont {Balewski}}, \bibinfo {author}
  {\bibfnamefont {B.}~\bibnamefont {Butscher}}, \bibinfo {author}
  {\bibfnamefont {H.~P.}\ \bibnamefont {B\"uchler}}, \ and\ \bibinfo {author}
  {\bibfnamefont {T.}~\bibnamefont {Pfau}},\ }\href
  {http://stacks.iop.org/0953-4075/45/i=11/a=113001} {\bibfield  {journal}
  {\bibinfo  {journal} {J.~Phys.~B: At.~Mol.~Opt.~Phys.}\ }\textbf {\bibinfo
  {volume} {45}},\ \bibinfo {pages} {113001} (\bibinfo {year}
  {2012})}\BibitemShut {NoStop}%
\bibitem [{\citenamefont {Balewski}\ \emph {et~al.}(2013)\citenamefont
  {Balewski}, \citenamefont {Krupp}, \citenamefont {Gaj}, \citenamefont
  {Peter}, \citenamefont {B\"uchler}, \citenamefont {L\"ow}, \citenamefont
  {Hofferberth},\ and\ \citenamefont {Pfau}}]{Balewski2013}%
  \BibitemOpen
  \bibfield  {author} {\bibinfo {author} {\bibfnamefont {J.~B.}\ \bibnamefont
  {Balewski}}, \bibinfo {author} {\bibfnamefont {A.~T.}\ \bibnamefont {Krupp}},
  \bibinfo {author} {\bibfnamefont {A.}~\bibnamefont {Gaj}}, \bibinfo {author}
  {\bibfnamefont {D.}~\bibnamefont {Peter}}, \bibinfo {author} {\bibfnamefont
  {H.~P.}\ \bibnamefont {B\"uchler}}, \bibinfo {author} {\bibfnamefont
  {R.}~\bibnamefont {L\"ow}}, \bibinfo {author} {\bibfnamefont
  {S.}~\bibnamefont {Hofferberth}}, \ and\ \bibinfo {author} {\bibfnamefont
  {T.}~\bibnamefont {Pfau}},\ }\href {\doibase 10.1038/nature12592} {\bibfield
  {journal} {\bibinfo  {journal} {Nature}\ }\textbf {\bibinfo {volume} {502}},\
  \bibinfo {pages} {664} (\bibinfo {year} {2013})}\BibitemShut {NoStop}%
\bibitem [{\citenamefont {Bahrim}\ \emph {et~al.}(2001)\citenamefont {Bahrim},
  \citenamefont {Thumm},\ and\ \citenamefont {Fabrikant}}]{Bahrim2001}%
  \BibitemOpen
  \bibfield  {author} {\bibinfo {author} {\bibfnamefont {C.}~\bibnamefont
  {Bahrim}}, \bibinfo {author} {\bibfnamefont {U.}~\bibnamefont {Thumm}}, \
  and\ \bibinfo {author} {\bibfnamefont {I.~I.}\ \bibnamefont {Fabrikant}},\
  }\href {http://stacks.iop.org/0953-4075/34/i=6/a=107} {\bibfield  {journal}
  {\bibinfo  {journal} {J. Phys. B: At. Mol. Opt. Phys.}\ }\textbf {\bibinfo
  {volume} {34}},\ \bibinfo {pages} {L195} (\bibinfo {year}
  {2001})}\BibitemShut {NoStop}%
\bibitem [{\citenamefont {Mayle}\ \emph {et~al.}(2012)\citenamefont {Mayle},
  \citenamefont {Rittenhouse}, \citenamefont {Schmelcher},\ and\ \citenamefont
  {Sadeghpour}}]{Mayle2012}%
  \BibitemOpen
  \bibfield  {author} {\bibinfo {author} {\bibfnamefont {M.}~\bibnamefont
  {Mayle}}, \bibinfo {author} {\bibfnamefont {S.~T.}\ \bibnamefont
  {Rittenhouse}}, \bibinfo {author} {\bibfnamefont {P.}~\bibnamefont
  {Schmelcher}}, \ and\ \bibinfo {author} {\bibfnamefont {H.~R.}\ \bibnamefont
  {Sadeghpour}},\ }\href {\doibase 10.1103/PhysRevA.85.052511} {\bibfield
  {journal} {\bibinfo  {journal} {Phys. Rev. A}\ }\textbf {\bibinfo {volume}
  {85}},\ \bibinfo {pages} {052511} (\bibinfo {year} {2012})}\BibitemShut
  {NoStop}%
\bibitem [{Rai()}]{Raithel}%
  \BibitemOpen
  \href@noop {} {}\bibinfo {note} {D. A. Anderson, S. A. Miller and G. Raithel,
  arXiv:1401.2477}\BibitemShut {NoStop}%
\end{thebibliography}
\end{document}